\documentclass[twocolumn]{aastex63}
\usepackage{newtxtext,newtxmath}
\usepackage[T1]{fontenc}
\usepackage{ae,aecompl}
\usepackage{graphicx}
\usepackage{amsmath}
\usepackage{amssymb}	
\usepackage{float}
\usepackage[]{hyperref}
\usepackage{xcolor}%New in the re-submitted version

\usepackage{booktabs}
\received{}
\revised{}
\accepted{}

\submitjournal{ApJ} 

\shorttitle{Positron Effects in M87* and Sgr A*}
\shortauthors{Emami et al.}

\graphicspath{{./}{figures/}}

\begin{document}

\title{Positron Effects on Polarized Images and Spectra from Jet and Accretion Flow Models of M87* and Sgr A*}

\correspondingauthor{Razieh Emami}
\email{razieh.emami$_{-}$meibody@cfa.harvard.edu}
  
\author{Razieh Emami}
\affiliation{Center for Astrophysics $\vert$ Harvard \& Smithsonian, 60 Garden Street, Cambridge, MA 02138, USA}

\author{Richard Anantua}
\affiliation{Center for Astrophysics $\vert$ Harvard \& Smithsonian, 60 Garden Street, Cambridge, MA 02138, USA}

\affiliation{Black Hole Initiative at Harvard, 20 Garden Street, Cambridge, MA 02138, USA
}

\affiliation{Center for Computational Astrophysics, Flatiron Institute, Simons Foundation, 162 Fifth Avenue, New York, NY 10010, USA
}

\author{Andrew  A. Chael}
\affiliation{Princeton Center for Theoretical Science, Jadwin Hall, Princeton University, Princeton, NJ 08544, USA}

\affiliation{NASA Hubble Fellowship Program Einstein Fellow}

\author{Abraham Loeb} 
\affiliation{Center for Astrophysics $\vert$ Harvard \& Smithsonian, 60 Garden Street, Cambridge, MA 02138, USA}

\begin{abstract}
We study the effects of including a nonzero positron-to-electron fraction in emitting plasma on the polarized SEDs and sub-millimeter images of jet and accretion flow models for near-horizon emission from M87* and Sgr A*. For M87*, we consider a semi-analytic fit to the force-free plasma regions of a general relativistic magnetohydrodynamic jet simulation which we populate with power-law leptons with a constant electron-to-magnetic pressure ratio. For Sgr A*, we consider a standard self-similar radiatively inefficient accretion flow where the emission is predominantly from thermal leptons with a small fraction in a power-law tail.  In both models, we fix the positron-to-electron ratio throughout the emission region. We generate polarized images and spectra from our models using the general-relativistic ray tracing and radiative transfer from \texttt{GRTRANS}.  We find that a substantial positron fraction reduces the circular polarization fraction at infrared and higher frequencies. However, in sub-millimeter images higher positron fractions increase polarization fractions due to strong effects of Faraday conversion.  We find a M87* jet model that best matches the available broadband total intensity and 230 GHz polarization data is a sub-equipartition, with positron fraction of $\simeq$ 10\%. We show that jet models with significant positron fractions do not satisfy the polarimetric constraints at 230 GHz from the Event Horizon Telescope (EHT). Sgr A* models show similar trends in their polarization fractions with increasing pair fraction. Both models suggest that resolved, polarized EHT images are useful to constrain the presence of pairs at 230 GHz emitting regions of  M87* and Sgr A*.
\end{abstract}


\section{Introduction}\label{sec:intro}
Supermassive black holes have an outsized influence on galactic dynamics --
catalyzing the radiatively inefficient inflow of billion-degree magnetized plasmas \citep[e.g.][]{Quataert2003} and the formation of energetic relativistic jets \citep[e.g.][]{Blandford1977}. The central supermassive black hole in the giant elliptical galaxy M87 launches an extragalactic jet which extends from several to tens of kiloparsecs across the EM spectrum. M87's jet has been studied for over a century from radio to $\gamma$-ray wavelengths 
\citep[e.g.][]{Curtis1918,Palmer67,Nagar2001,Doeleman2012,Whysong2004,Madrid2009,DiMatteo2003,Aharonian2006,MWL2021}.
Very-Long-Baseline Interferometry (VLBI) images at radio and sub-millimeter wavelengths show a dynamical, polarized, limb-brightened forward jet and counter jet in the central parsecs of M87 \citep[e.g.][]{Junor1999,Asada2012,Hada2016,Walker2018,Kim2018,Kravchenko2020}. The jets terminate in a bright radio core that corresponds to the location of the central black hole at frequencies $\gtrsim 86$ GHz \citep{Hada2011}.

Recently, the Event Horizon Telescope (EHT) observed M87* (the central black hole in M87) at 230 GHz and produced the first resolved images of 
emission within $\lesssim 10$ gravitational radii $r_{\rm g}$ of a black hole\footnote{Throughout the text, we use the gravitational radius $r_{\rm g}=GM/c^2$ and gravitational timescale $t_{\rm g}=GM/c^3$ for a black hole of mass $M$.}  \citep{EHTI,Akiyama:2019brx,Akiyama:2019sww,Akiyama:2019bqs,Akiyama:2019fyp,Akiyama:2019eap}. The EHT images show a ring of synchrotron emission 
approximately $40\,\mu$as in diameter, consistent with the expected lensed photon orbit size for a black hole of mass $6.5\times10^9M_\odot$ \citep{Akiyama:2019eap}. The emission ring in the EHT images is brighter in its southern half, consistent with the effects of Doppler beaming from emitting plasma rapidly co-rotating with the black hole \citep{Akiyama:2019fyp, Wong2020}. 

Notably, the near-horizon synchrotron emission from M87* is linearly and circularly polarized \citep{Akiyama:2021eap,Akiyama:2021fyp,Goddi2021}. 
The 230 GHz linear polarization resolved by the EHT has a 
relatively low resolved linear polarization fraction $|m|\sim20\%$ at the $\approx20\,\mu$as scale of the EHT beam and a 
predominantly azimuthal, spiral pattern of electric vector position angles (EVPAs) around the emission ring \citep{Akiyama:2021eap}. The circular polarization fraction of the 230 GHz ring is low, $\lesssim 0.4$\% \citep{Goddi2021}. These polarization data are much more constraining on physical models of M87* than total intensity data alone. While \citet{Akiyama:2019fyp} found that many general relativistic magnetohydrodynamic (GRMHD) simulations of M87* could produce images consistent with the observed emission ring, \citet{Akiyama:2021fyp} found that only a few simulation images fit the polarimetric data; these images were all from simulations with strongly ordered poloidal magnetic fields in the magnetically arrested disk (MAD) state  \citep[e.g.][]{Narayan2003}. 

Despite these groundbreaking observational results, persistent questions on the fundamental nature of jets connected to black holes remain, including: ``What physical mechanisms accelerate the emitting particles and what is their energy distribution?"
and
``Is the particle composition of relativistic jets near supermassive black holes ionic, leptonic, or mixed?"
With regard to the second question, most analytic and simulation models of the near-horizon emission from low-luminosity AGN (LLAGN) like M87* 
and the galactic center black hole Sagittarius A* (Sgr A*)
assume a purely ionic plasma when comparing observables to data. For instance, the library of GRMHD simulations compared to the resolved polarimetric images of M87* in  \citet{Akiyama:2021fyp} all assumed positron emission does not contribute to the observed image. While some groundbreaking studies have estimated the positron fraction around M87* and Sgr A* from different models of pair production \citep[e.g.][]{Moscibrodzka2011,Broderick2015,Wong2021}, their implications for resolved images of the sources have not been studied in detail.
In this work, be begin to assess the effects of a non-zero pair fraction on submm images and broadband spectral energy distributions (SEDs) of M87* and Sgr A* using 
semi-analytic models for both sources. 
GRMHD simulations of an electron-proton plasma are the primary modern theoretical tools for understanding the physics of jet/accretion flow/black hole (JAB).  Numerical GRMHD codes such as {\tt HARM} \citep{Gammie2003}, {\tt Athena++} \citep{White2016}, and {\tt BHAC} \citep{Porth2017}
can accurately solve for fluid variables 
in most regions of JAB systems. Such codes have served as a bridge between the microphysics of particle acceleration, emission (including the conversion of magnetic to particle energy) and discrete observational features in AGN, e.g., Sgr A* \citep[e.g.][]{Moscibrodzka2009,Dexter2010,Shcherbakov2012,Chan2015, Ressler2016,2017ApJ...837..180G,Anantua2020b,Chael2018,Dexter2020} and M87* \citep[e.g.][]{Dexter2012,Moscibrodzka2018, Anantua2018,Ryan2018, Chael2019, Davelaar2019}, but all images produced from GRMHD simulations to date have assumed a plasma free of electron-positron pairs.  
GRMHD simulations reliably solve the flows in the Kerr metric in relatively high-density disk regions, but they 
have difficulty in reliably solving for the plasma temperature and dynamics in the low density, magnetized jet -- particularly where the magnetization ($\sigma=\frac{\mathrm{magnetic \ energy \ density}}{\mathrm{plasma \ density}}$) 
exceeds unity. The magnetic field is more accurately evolved in jet regions in GRMHD simulations.
If we wish to produce emission models from near-horizon jets in M87* or Sgr A*, it may be useful to 
begin by using GRMHD magnetic field results as the basis for self-similar/semi-analytic models of the emission region. 

While they cannot capture the dynamics and turbulence of GRMHD simulation images of M87* and Sgr A*, semi-analytic models allow for a wider scan of the parameter space, can extend self-consistently to large radii, and can avoid pathologies in fluid variables in the jet region. The foundational semi-analytic model of synchrotron emission from relativistic jets is from \citet{Blandford1979},
where a helical magnetic field near equipartition with the radiating particle population naturally accounts for the radio spectral slope. Applications of semi-analytic models of jet emission to M87* include \citep[e.g.][]{Reynolds1996,BroderickLoeb2009,Prieto2016}, and the flat radio Sgr A* SED can also be naturally explained by a jet model  \citep[e.g.][]{FalckeMarkoff}.
A plausible alternative to a jet origin for observed near-horizon submm emission in M87* and Sgr A* is that the emission originates in a geometrically thick equatorial accretion flow.  
In the regime in which gas accreting towards a black hole is unable to radiate efficiently,  \cite{Narayan1994} produced self-similar advection-dominated accretion flow (ADAF) solutions with positive Bernoulli constant (enabling them to  become outflows if they escape accreting onto the black hole). 
ADAF and similar radiatively inefficient accretion flow (RIAF) models for Sgr A* 
can fit the observed SED and constraints on the intrinsic submm size \citep[e.g.][]{NarayanYi1995,Quataert1999,Ozel2000,Broderick2006a}. They predict a highly sub-Eddington accretion rate with a quiescent X-ray spectrum dominated by bremsstrahlung near the Bondi radius, sub-mm synchrotron emission from a mixed thermal- and nonthermal electron distribution near the black hole, and
and near-infrared and X-ray flares originating  $\lesssim$ 10$r_g$ from the black hole \citep[e.g][]{Quataert2003}. In addition to pure ADAF or pure jet models, hybrid jet+ADAF models where different emitting regions contribute to different parts of the observed M87* or Sgr A* SED have also been compared to observations \citep[e.g.][]{Yuan2002,Feng2016}.

In this paper, we provide examples of semi-analytic models of JAB systems tuned to simulate the emission from M87* and Sgr A* while including the effects of mixed pair and ionic plasmas. Positrons can be created through various channels in JAB systems, including photon annihilation in the disk and jet funnel walls, and spark gap acceleration \citep{Moscibrodzka2011, Broderick2015}. We utilize the general relativistic ray tracing and radiative transfer (GRRT) code {\tt GRTRANS} \citep{Dexter2016}, which computes synchrotron Stokes maps from observer-to-source rays in the Kerr metric through an emitting plasma around a Kerr black hole. 
We focus on images from semi-analytic models of M87* and Sgr A*. We analyze the effects of pairs on GRMHD simulation images by replacing some proton positive charge carriers with positrons in the post-processing step of our radiative transfer. 
Incorporating the effects of pairs in radiative transfer of physical models of M87* and Sgr A* is a key step towards understanding how EHT and broadband measurements may constrain the composition and nature of plasma in such sources. 

This paper is organized as follows:  Section~\ref{sec:observations} describes recent observations of M87* and Sgr A* that we make use of in comparing our models to data. Section~\ref{sec:estimates} summarizes two recent models for positron production in the context of JAB systems and their implications for the positron-to-electron fraction $f_{\rm pos}$ in M87* and Sgr A*. Section~\ref{sec:models} describes the semi-analytic jet and disk models we use. Section~\ref{sec:modelspace} describes our parameterization for incorporating the effects of a mixed ionic and leptonic plasma on the images from these models. Sections~\ref{sec:Model-search} and ~\ref{sec:Results} present images and polarized spectra from these models and compare to observations.  
Section ~\ref{sec:summary} concludes and proposes future directions.  

\section{Observations of M87 and Sgr A*}
\label{sec:observations}

Below we introduce observational conventions before we describe different observations for M87* and Sgr A*. 

\subsection{Conventions}
In this paper, we produce polarized images from semi-analytic models at frequencies for which we can compare to VLBI data from M87* and Sgr A*. We first establish our polarimetric conventions. Synchrotron radiation from a single emitter is elliptically polarized. Denoting the ellipse orientation angle $\psi$ and the ellipticity angle $\chi$ for an incoming electromagnetic wave with electric vector projection $\vec{E}_0$ on the observer plane, we use standard definitions of the Stokes parameters $I, Q, U$ and $V$:
\begin{align}
    I &= E_0^2,  \\
    \tan 2\psi &= \frac{U}{Q}, \\
    \sin 2\chi &= \frac{V}{I}.
\end{align}
We measure the electric vector position angle (EVPA) $\psi$ in degrees East of North. 

Throughout this work, we consider two different types of fractional polarization: an unresolved fractional polarization and an average, resolved fractional polarization. For both M87* and Sgr A*, most of our information on the polarimetric properties of the near-horizon accretion flow and jet launching region is unresolved. To compare models against the data, we compute the unresolved, net polarization fractions $|m|_{\rm net}$ and $|v|_{\rm net}$:
\begin{align}
\label{eq:polfracs1}
|m|_{\mathrm{net}} \equiv & \frac{\sqrt{\overline{Q}^2 + \overline{U}^2}}{\overline{I}}\\
\label{eq:polfracs2}
|v|_{\mathrm{net}} \equiv & \frac{\overline{|V|}}{\overline{I}},
\end{align} 
where bars indicate averages over pixels in a source model image. For M87*, for the first time, EHT results in \citep{Akiyama:2021eap} constrain the resolved linear polarization near the horizon. They define the average resolved polarization fraction as
\begin{equation}
\label{eq:polfracs3}
    \langle |m| \rangle \equiv \frac{\overline{\sqrt{Q^2 + U^2}}}{\overline{I}}.
\end{equation}
The average resolved polarization fraction $\langle |m| \rangle$ depends on the observation resolution, while the unresolved fractions $|m|_{\rm net}$ and $|v|_{\rm net}$ do not. In comparing results from 230 GHz theoretical images, we simulate the effects of limited observational resolution by blurring the image with a circular Gaussian filter at the EHT resolution of 20 $\mu$as. 

We note that in addition to $|v|_{\rm net}$, $|m|_{\rm net}$ and $\langle |m| \rangle$, in comparing GRMHD simulation images to EHT polarimetric data, \citet{Akiyama:2021fyp} also made use of the second azimuthal Fourier model of the complex linear polarization $\beta_2$ \citep{PWP2020}. Here we do not consider this constraint on the model images of M87*, as it is predominantly set by the field geometry in our model, which we do not vary as a free parameter. Nonetheless, we note that the EVPA patterns we recover in our jet model for M87* are predominantly azimuthal and are somewhat qualitatively similar to the observed M87* image. 

\subsection{M87 Observations}

The giant elliptical galaxy M87 hosts our first- and best-observed jet emanating from a supermassive black hole, as well as the first resolved image of a black hole's near-horizon environment. This section briefly summarizes some recent observations of M87* applicable to this work. The jet from M87 has been observed from radio to $\gamma$ ray frequencies; a comprehensive SED constructed as part of simultaneous and near-simultaneous observations to the 2017 EHT campaign was recently published in \citet{MWL2021}. 

VLBI images at frequencies between $24-86$ GHz \citep[e.g.]{Walker2018,Hada2016,Kim2018,Kravchenko2020} show the jet structure within the central $\sim 1$pc. The jet is dynamical and polarized, with a wide opening angle $\sim55^\circ$ at 43 GHz. \citet{Hada2016} fits the opening angle with distance from the black hole as $s\sim z^{0.56}$. The inclination angle toward the line of sight is $\theta_o\approx 17-30^\circ$ \citep{Mertens2016}. The forward jet is limb-brightened, and a counter-jet is visible in some images. There is no core-shift with frequency above $\approx 86 $ GHz, which suggests that the radio core at these frequencies is becoming optically thin and is fixed at the black hole's location \citep{Hada2011}. 

Polarized images of the jet generally show low polarization fractions near the core with stronger polarization fractions along the jet. In 43 GHz images from \citet{Walker2018}, the polarization fraction peaks at $4\%$ in these images along the edge of the jet. The EVPA structure in the 43 GHz images suggests that the jet magnetic field is predominantly toroidal. \citet{Hada2016} found the linear polarization fraction at 86 GHz is of the order $4\%$ near the core, but features further down the jet rise to a level $\sim20\%$; they also infer a poloidal field structure in the jet. \citet{Kravchenko2020} observed the jet from 24-43 GHz and found that the global polarization pattern was stable over 11 years of observations, with some variability on top of the global pattern on $\sim$ month timescales. The polarization pattern again indicates a predominantly toroidal jet magnetic field near the jet launching site. They estimate a rotation measure (RM) toward the core of $\sim10^4$ rad m$^{-2}$ between 24 and 86 GHz; they interpret this RM as due to an external screen, possibly in a non-relativistic wind outside of the jet. 
As part of the 2017 EHT campaign, \citet{Goddi2021} performed observations of M87* with ALMA at a central frequency $\nu=221$ GHz; these observations do not resolve the inner jet or core, but they were used to estimate the unresolved linear and circular polarization $|m|_{\rm net} \approx$ 2.7\%, $|v|_{\rm net}\leq 0.4 $\%. \citep{Goddi2021} also measured the core RM at 221 GHz; they found a rotation measure that is time variable with a maximum magnitude of $10^5$ rad m$^{-2}$. This RM varies in magnitude and sign by order $\sim$1 over the week of observations in 2017. While substantial RM can be produced close to or internal to the submm emission region \citet[e.g.][]{Ricarte2020}, it is still uncertain whether the observed RM occurs close to or far away from the emission region. In this paper, we only consider the effects of RM internal to our models and do not impose an external Faraday screen. 

The EHT has released 230 GHz M87* image data from its 2017 observing campaign in eight papers \citep{Akiyama:2019cqa,Akiyama:2019brx,Akiyama:2019sww,Akiyama:2019bqs,Akiyama:2019fyp,Akiyama:2019eap,Akiyama:2021eap,Akiyama:2021fyp}.
The 230 GHz images feature a ring of emission of mean radius $42\mu$as $\pm 3 \mu$as that is brighter in the South than the North  \citep{Akiyama:2019cqa}. 
Proto-EHT observations with three stations from 2009-2013 also show evidence for a ring morphology with a consistent radius \citep{Wielgus2020}.
The 2017 EHT 230 GHz polarized intensity maps of M87 \citep{Akiyama:2021eap,Akiyama:2021fyp} show azimuthal EVPA patterns and a low level of resolved linear polarization ($\langle|m|\rangle \lesssim 11\%$). 
The only GRMHD simulations from the EHT simulation library that pass polarimetric constraints derived from these images are of Magnetically Arrested Disk (MAD) systems, which feature dynamically important poloidal magnetic fields. 
These passing simulations prefer near-horizon magnetic fields of 7-30 G, densities in a range $10^{4-5}$ cm$^{-3}$ and electron temperatures in a range from $(5-35)\times10^{10}$ K. These results imply a value of plasma-$\beta_{\rm e}$ (where $\beta_{\rm e}=\frac{\rm plasma \ pressure}{\rm magnetic \ pressure}$) in the range $10^{-3} \lesssim \beta_{\rm e} \lesssim 1$ in the 230 GHz emission region \citep{Akiyama:2021fyp}.

We take the full SED of M87 from radio to X-ray frequencies from the table compiled by \cite{Prieto2016}, focusing on their results on scales less than of 0.4 as (32 pc).\footnote{Note that a more recent and complete M87* SED is presented in \citet{MWL2021}. Because we calibrated our M87* images to the 230 GHz flux density from \citet{Akiyama2015}, which is a factor of 2 larger than the 2017 value observed by the EHT, we use the \citet{Prieto2016} compiled SED, which includes this value, for consistency.} \citet{Prieto2016} compile observations previously published at various frequencies in \citet{Aharonian2006,DiMatteo2003,Perlman2001,Whysong2004,Nagar2001,Doeleman2012,Lonsdale1998,Lee2008,Junor1995,Morabito1986,Morabito1988}. In addition, we make use of updated estimates of the core compact flux density first compiled in from data originally published in \citet{Lister2018,Hada2017,Walker2018,Kim2018,Doeleman2012,Akiyama2015}.
The total flux densities from these VLBI observations (from 15-230 GHz) were re-measured and compiled in \citet{Chael2019} using a consistent strategy across all data sets.

A synchrotron peak in M87's SED occurs at $\nu_\mathrm{syn}=10^{11.5}$ Hz; a secondary peak, which may be interpreted as due to synchrotron self-Compton or from a secondary nonthermal synchrotron-emitting population, occurs at $\nu_\mathrm{SSC}=10^{14.5}$ Hz.  Between the infrared and X-ray bands, both the observed quiescent and loud states have spectral flux density $S_{\nu} = \nu^{-\alpha}$ with a slope $\alpha \sim 1.1$. If we assume this slope originates from synchrotron emission from power-law electrons, we can estimate the power-law index of the emitting distribution as $p \approx 3.5.$

The M87* core flux density we use to normalize our 230 GHz images throughout this paper is 1.5 Jy \citep{Akiyama2015}. Note that in the 2017 EHT observations, the core flux at 230 GHz had decreased to 0.6 Jy.  While we compare our M87* jet models to the full Stokes $I$ SED, note that when considering polarimetric constrains on the models in this paper, we only consider the 2017 EHT polarimetric constraints and do not apply constraints from polarimetric observations at longer wavelengths. This is because \citet{Akiyama:2021fyp} provide a straightforward framework for comparison in terms of $|m|_{\rm net}$, $|\langle |m| \rangle$, and $|v|_{\rm net}$, since we are most interested in modeling the 230 GHz source structure, and as we find the 230 GHz polarimetric constraints to be sufficiently constraining (in fact, they reject most of our models.) In a future work we will consider a wider range of polarimetric constraints from multi-wavelength data. 
 
\subsection{Sgr A* Observations}\label{sgrA-observation}
Sgr A* is the most extensively studied accreting supermassive black hole. Its proximity ($D=8.178\pm22$ kpc) and mass of $4.154\pm0.014 \times 10^6 M_\odot$ \citep{Gravity2019} make it the largest angular size black hole known. However, it is situated in a bath of dust and gas which-- along with its intrinsic variability-- make it particularly challenging to observe and to model. Sgr A* has been observed from radio to gamma-ray wavelengths. At near-infrared and X-ray wavelengths, the continual flaring behavior of Sgr A* has attracted particular interest, as it promises to help elucidate the rapid dynamics and microphysical processes of hot plasma near the event horizon 
\citep[e.g.][]{Yusef06,Gillessen2006,Marrone2008,Neilsen2013,Ponti2017,Gravity2018a,Do2019}. 
Sgr A*'s emitting region at submm wavelengths is compact. At frequencies $<230$ GHz, strong interstellar scattering prevents the direct imaging of Sgr A*'s intrinsic structure from VLBI observations. Instead, most observations constrain the size of the scattering kernel that blurs the source image \citep[e.g.][]{Johnson2018}.
\citet{Issaoun2019} produced the first directly resolved VLBI images of Sgr A* at 86 GHz from GMVA+ALMA data that constrain the intrinsic structure of the source. Their 86 GHz estimates for Sgr A*'s axial sizes assuming an elliptical profile are ($100\pm 18) \mu$as, or ($20.2\pm 3.6)r_g$. The comparison of these size constraints in \citet{Issaoun2019} to theoretical models from GRMHD simulations ruled out jet-dominated emission models at 86 GHz unless the jet is nearly face-on. 

Sgr A* has been observed by ``proto-EHT" arrays (with 3-4 sites) from 2007-2013. The intrinsic source size at 230 GHz was first measured by \citet{Doeleman2008} using proto-EHT data as $\approx40\mu$as; this size was confirmed with updated proto-EHT measurements including longer baselines to APEX in Chile by \citet{Lu2018}.
De-scattered image moments were  calculated from 2013 proto-EHT observations by \citet{Johnson2018}, who report an major axis angular width $9.9r_g<\theta_\mathrm{maj}<12.7r_g$ \citep{Johnson2018}. \citep{Broderick2006a} modeled proto-EHT observations of Sgr A* from 2007 to 2013 modeled with semi-analytic RIAF models find $\theta\approx60^\circ$ and a low black hole spin. 

The polarized emission from Sgr A* has also been extensively studied at sub-mm wavelengths \citep[e.g.][]{Bower1999,Marrone2006,Marrone2008,Yusef2007,Munoz2012,Johnson2015,Bower2018}. 
At 230 GHz, Sgr A* is significantly linearly polarized with mean polarization fractions ranging from $|v|_{\rm net} \approx 4-8$\% \citep{Bower2018}. Like the total intensity at submm-wavelengths, the linear polarization exhibits variability on $\sim$hour timescales \citep{Marrone2006,Yusef2007}. \citet{Bower2018} also provides the latest measure of the rotation measure at 230 GHz $RM\approx-5\times10^5$ rad m$^{-2}$, consistent with its value measured over 20 years \citep[e.g.][]{Marrone2007}. The measured RM can be used to infer an accretion rate onto Sgr A* of $\approx 10^{-8} M_\odot $ yr$^{-1}$ (though note that the accretion rate measured via the RM can produce inaccurate results for face-on systems \citep{Ricarte2020}). \citet{Munoz2012} found that Sgr A* is also significantly circularly polarized at 230-345 GHz, with  $|v|_{\rm net} \approx 1-2$\% GHz. The (negative) sign of $V$ is consistent with that measured at longer wavelengths, suggesting that it arises from Faraday conversion in the accretion flow with a stable magnetic field orientation. \citet{Johnson2015} provided the first \emph{resolved} look at Sgr A*'s polarimetric structure from proto-EHT observations; they found that the EVPA pattern on event-horizon scales (and hence the magnetic field configuration) exhibits an intermediate degree of spatial order with coherent structures on scales of $\sim10r_{\rm g}$. The average 230 GHz resolved polarization fraction from these measurements is $10 \%  \leq  \langle |m| \rangle  \leq   30 \%$.

Intriguingly, GRAVITY, a near-infrared interferometer, has observed Sgr A* \citep{Gravity2018a} at 2.2 microns and detected circular motions that potentially arise from semi-coherent ``hotspots" orbiting in a near face-on accretion flow between $6r_g$ and $10r_g$. These circular motions arise during infrared flares from Sgr A*; in addition to signatures of a circular motion from interferometric data. GRAVITY observed ``loops" in the near-infrared EVPA which suggests that the emitting hot spots are situated in a vertical magnetic field. Similar ``loops" have been seen in submm observations \citep{Marrone2006}. These hotspots may correspond to the formation of magnetic flux tube bundles, or plasmoids, which seen in GRMHD simulations with resistivity enabling plasmoid formation through magnetic reconnection \citep{Ripperda2020, ball2020plasmoid}. 

The Stokes $I$ Sgr A* spectrum we use is compiled in the Appendix of \cite{Ressler2016}; it spans from $10^{10.5}$ Hz radio to $10^{10.5}$ Hz X-rays. It includes radio and millimeter data
from \citet{Falcke1998,An2005,Bower2015,Liu2016a,Liu2016b,Doeleman2008} and \citet{Johnson2015}. Infrared data were taken from \citet{Cotera99,Genzel99,Genzel03,Schodel07,Schodel11} and \citet{Witzel2012}.
X-ray data over the range 2-10 keV were taken from \citet{Neilsen2013,Baganoff03}. In this paper, we normalize our 230 GHz Sgr A* images so that the total flux is 3.5 Jy from \cite{Bower2015}. Note that earlier, proto-EHT visibility amplitude versus baseline length measurements gave a value of 2.4 Jy \citep{Doeleman2008}.

\section{Pair Production in Accretion/Jet systems}
\label{sec:estimates}

Particles near the supermassive black hole event horizon in LLAGN systems like M87* and Sgr A* are relativistically hot, with temperatures exceeding $10^{10}$ K $\approx 1$ MeV. The near-horizon environment thus features many particles with energies above the pair creation threshold. In principle, electron-positron pairs can be produced by several different channels including: particle-particle collisions ($e^-p$ or $e^-e^-$), photon-particle collisions (the Bethe-Heitler process, $\gamma e^-$ or $\gamma p$) and photon-photon collisions (the Breit-Wheeler process, $\gamma\gamma$). In practice, because plasma density is extremely low in LLAGN (with mean free paths exceeding the system size) and because photon-photon collisions have the greatest cross section near the pair-production threshold \citep[e.g.][]{Stepney1983}, photon-photon collisions dominate pair production in astrophysical models of these sources \citep[see also][]{Moscibrodzka2011}. That is, in this work we only consider models where pair production arises from: 
\begin{equation}
    \gamma + \gamma\to e^- + e^+.
\end{equation}
Whether or not there is a substantial pair population in the near-horizon region of low luminosity AGN (LLAGN) like Sgr A* and M87* remains an open question. On the one hand, in models where an accretion disk produces most of the observed luminosity, a large pair fraction is disfavored. For instance, in M87* the temperature of the synchrotron radiating particles at 1 mm is likely in the range $5-30$ MeV \citep{Akiyama:2021fyp}. As a first estimate, we can assume this temperature is close to the disk virial temperature $T_{\rm vir}$, which for a disk with particles of mean mass $\bar{m}$ and at the distance $r$ from the gravitating mass $M$ is set by
\begin{equation}
\label{eq:vir}
k_BT_\mathrm{vir}=\frac{1}{3}\frac{GM\bar{m}}{r}.
\end{equation}
Equation~\ref{eq:vir} implies that at a radius e.g. $r\approx 5\,r_{\rm g}$, $k_BT_{\rm vir}\approx 60$ MeV, within an order of magnitude of the observed temperature. However, 
since the electron mass is a fraction 1/1,836 of the proton mass if we replace each proton with a positron in an accretion disk, the virial temperature would be two orders of magnitude below the observed value. 
While a fully pair-dominated disk is thus unlikely, we could instead have a scenario where a small population of pairs is produced in the disk-jet corona from $\gamma\gamma$
\citep{Laurent2018,Moscibrodzka2011}. This pair population could be subdominant in determining the overall structure of the observed image at millimeter wavelengths, but it could still affect the polarized source structure.  
In contrast, the scenario may be quite different if most of the emission at 1 mm originates from nonthermal particles in a relativistic jet with substantial Doppler beaming and relativistic aberration. \citet{Reynolds1996} used a \citet{Blandford1979} model of nonthermal emission in a magnetized jet to put constraints on the emitting particle density $n$ and magnetic field $B$ of M87's jet emission at 5 GHz. They find, if they demand the observed 5 GHz core is just becoming optically thick,  the observed surface brightness and total jet power prefer an emitting particle distribution in the jet dominated by pairs. Similarly, in the broader class of Blazar jets, \citet{Ghisellini2012} finds that the inner regions of observed Blazar jets can efficiently produce positrons with densities an order of magnitude greater than the ion density as long as the gamma-ray luminosity $\gtrsim10^{44}$ erg s$^{-1}$. We note, however that the observed gamma-ray luminosity of M87*'s core is $\sim10^{42}$ erg s$^{-1}$ \citep{MWL2021}.

In this section, we review two classes of models that have been used to estimate pair production efficiencies in near-horizon accretion flow and jets models of Sgr A* and M87*; 
gap acceleration models, where pairs are produced from gamma-ray photons originating in an inverse-Compton cascade from seed particles accelerated by unscreened electric fields in the jet,
and ``pair-drizzle" scenarios, where pairs are produced directly from photons emitted by the bulk accretion flow and upscattered by inverse Compton collisions. The models we summarize here provide some motivations for the range of positron-to-electron fractions we explore in semi-analytic models of M87* and Sgr A* discussed later in the paper. However, note that in our models in the remainder of this work, we are intentionally agnostic about the physical source of pairs; instead, we focus on the observational consequences for polarized emission, if a fraction of pairs is present from any physical production channel. Note in particular that while both of the production models we discuss in this section produce non-negligible pair densities for M87* (with positron-to-electron fractions from $10^{-3} - 1)$, both models produce near-zero positron fractions for Sgr A*. Nonetheless, in Section~\ref{sec:modelspace} we consider RIAF models for Sgr A* and investigate their images under the same sampling of $f_{\rm pos}$ as we do for M87*. 

\subsection{Gap acceleration and pair cascade}
Gap acceleration occurs in magnetospheres where there are unscreened electric fields ($\mathbf{E}\cdot\mathbf{B}\neq0$), so particles directly accelerate particles to high energies \citep{GJ1969}. These accelerated particles inverse Compton scatter on ambient photons, which then collide with other photons to produce pairs in a cascade. This is the canonical mechanism for supplying pulsar magnetospheres with plasma; it has been applied to black hole magnetospheres in analytic models by \citep[e.g.][]{Blandford1977,Beskin1992,Hirotani1998,Vincent2010, Broderick2015}. Recently, this mechanism has been applied to full general relativistic particle in cell simulations of black hole jets by \citep{Chen2018,Levinson2018}. Here we summarize the results of \citet{Broderick2015} and their implications for potential pair fractions in M87* and Sgr A*. 

\citet{Broderick2015} assume that particles are accelerated to initiate a pair cascade at the stagnation surface that divides the outer region of the jet where matter accelerates outward from the region where it falls into the black hole. 
The charge-starved region around such a surface allows the build-up of an unscreened, coherent electric field \citep[][equation 1]{Broderick2015}
\begin{equation}
    E\approx \frac{s\Omega_B B}{c},
\end{equation}
where $s$ is the cylindrical radius and $\Omega_B$ is the angular velocity of a magnetic field line (assumed to be some fraction of the horizon angular velocity $\Omega_{\rm H}=ac/2r_{\rm H}$, where $a$ is the black hole spin and $r_{\rm H}$ is the horizon radius). This unscreened electric field forms a ``spark gap" that accelerates leptons to to high energies with energy $\gamma_{\rm max}m_{\rm e}c^2 \approx e E \ell$, where $\ell$ is the length scale over which particles are accelerated. After reaching their maximum Lorentz factor $\gamma_{\rm max}$, accelerated electrons cool by inverse Compton scattering off of ambient seed photons. If the emitted photons have sufficient energy they will produce pairs in an inverse Compton cascade. The limiting Lorentz factor below which the cascade halts, $\gamma_{\gamma \gamma}$, is set by the density of photons with energies above the pair production threshold: $\epsilon_{\rm th}\gtrsim 4m_{\rm c}c^2/\gamma_{\rm max}$.  
The total number of leptons produced in the post-gap cascade is then enhanced by a factor $\gamma_{\rm max}/\gamma_{\gamma\gamma}$ from its value at the top of the gap.

In practice, electrons are accelerated to a distance $\ell \approx \ell_{\rm IC}$ set by the inverse-Compton cooling length: 
\begin{equation}
    \ell_{\rm IC} = \frac{3m_{\rm e}c^2}{4\sigma_{\rm T}u_s\gamma_{\rm max}}, 
\end{equation}
where $u_s$ is the energy density of seed photons and $\sigma_{\rm T}$ is the Thomson cross-section \citep[][Equation 6]{Broderick2015}. 
For M87* and Sgr A*, they assume $u_s$ is the same as the the peak observed flux density \citep[][Equation 5]{Broderick2015},
\begin{equation}
    u_s = \frac{3 L_s}{4 \pi r_s^2 c}
\end{equation}
where $L_s$ is the observed bolometric luminosity and $r_s\approx 10r_{\rm g}$. 
For M87*, $L_s\approx10^{41}$ erg s$^{-1}$, so $u_s\approx 10^{-2}$ erg cm$^{-3}$. Assuming a black hole spin $a=0.9$ and requiring the jet power $P\approx10^{44}$ erg s$^{-1}$ to arise from the BZ mechanism, \citet{Broderick2015} calculate a field strength $B\approx35$ G at the stagnation surface $r\approx 10r_{\rm g}$.  Using $\ell_{\rm IC}$ as the cutoff acceleration length, they find $\gamma_{\rm max}\approx 1.6\times10^9$, so the (lab frame) seed photon pair production threshold is at $\epsilon_{\rm th}\approx 1.2$ meV. Critically, this is slightly below the spectral break, so the density of available photons for the post-gap cascade is maximized. As a result, while \citet{Broderick2015} find a pair density of only $n_g\approx2.2\times10^{-2}$ cm$^{-3}$ at the top of the gap, for M87*, this initial density is enhanced by a factor $\gamma_{\rm max}/ \gamma_{\gamma\gamma} \approx 670$ in the post-gap cascade to a final lepton density of $n_{\rm lep} \approx 15$ cm$^{-3}$. \citep{Broderick2015} also provide an estimate for the number density of emitters from observations of M87*, assuming $dn/d\gamma\sim\gamma^{-3.4}$, a source size of $40\,\mu$as, a 230 GHz flux of 1 Jy, and a magnetic field $B\approx 35$ G; this value is also  $n_{\rm emit}\sim15$ cm$^{-3}. $\footnote{Note that the \citet{Broderick2015} estimate of the number density of 230 GHz emitters is substantially lower than the estimate of $10^{4-5}$ $cm^{-3}$ from \citet{Akiyama:2021fyp}. his is because they assume a nonthermal energy distribution beginning at $\gamma_{\rm min}\approx 100$, while the EHT simulation modeling work assumes thermal particle distributions with lower mean energies.} Since $n_{\rm lep} = 2n_{e^+}$ and $n_{\rm emit} = n_{e^-}+n_{e^+}$, for the ratio of electrons to emitters we have:
\begin{equation}
    f_{\rm pos} = \frac{n_{e^+}}{n_{e^-}} = \frac{n_{\rm lep}}{2n_{\rm emit} - n_{\rm lep}},
\end{equation}
which gives an estimate of:
\begin{equation}
    f_{\rm pos} \approx 1 \;\;\; (\mathrm{M87^*}).
\end{equation}
In contrast, for Sgr A*, \citet{Broderick2015} use a luminosity $L\approx10^{36}$ erg s$^{-1}$, estimate a magnetic field of $B\approx 10^{2-3}$ G, and an initial seed photon energy density $u_s\approx 1.5\times10^{-2}$ erg cm$^{-3}$. The maximum Lorentz factor from gap acceleration is then $\gamma_{\rm max}\approx 7.6 \times 10^8$. As a result, the threshold energy for inverse Compton scattering is at a higher energies than in M87* $\epsilon_{\rm th}\approx 2.7$ meV. This is significantly above Sgr A*'s spectral break at $\approx 1$ meV. Consequently, there are substantially fewer photons available for the pair cascade in Sgr A* than in M87*. Therefore, the mean free path of pair production $\ell_{\gamma\gamma}\approx 10r_{\rm g}$. Because $\ell_{\gamma \gamma}$ is on the scale of the system size, there is no efficient pair cascade after the gap acceleration, $n_{\rm lep} = n_g \approx 2.5\times10^{-3}$ cm$^{-3}$. Similar to M87*, \citet{Broderick2015} infer a total number density of emitters from fitting the 230 GHz source size and flux density of 3 Jy to be $n_{\rm emit} \approx 10^6$ cm$^{-3}$. As a result, the positron-to-electron ratio can be inferred: 
\begin{equation}
    f_{\rm pos} \approx 10^{-9} \;\;\; (\mathrm{Sgr \, A^*}).
\end{equation}
Thus, the \citet{Broderick2015} pair cascade model predicts two qualitatively distinct values of $f_{\rm e}$ for Sgr A* and M87*. Because the post-gap cascade in M87* is efficient due to the higher density of seed photons above the energy threshold, $f_{\rm pos}\approx 1$ and the plasma in their model is assumed to be pair-dominated. In contrast, for Sgr A* the post-gap cascade is not efficient and $f_{\rm pos}\approx 0.$

 \subsection{``Pair Drizzle"}
Positrons in jets may also originate from annihilating high-energy photons produced by incoherent structures in the accretion flow, rather than by photons produced in the coherent gap regions in a cascade \citep{Moscibrodzka2011,Wong2021}. In these ``pair drizzle" models, high-energy photons above the pair-production threshold are initially produced in an accretion flow or jet wall by synchrotron self-Compton. Because the photons produced by inverse Compton scattering off of typical thermal accretion flow particles are less energetic than those produced in a pair cascade in a spark-gap scenario, these models are less efficient at producing pairs (typically, each photon accelerated by inverse Compton can only produce one pair).

\citet{Moscibrodzka2011} performed the detailed calculation of pair-production efficiency from a GRMHD simulation tuned to both Sgr A* and M87*. After determining the simulation accretion rate that produces the correct flux density at 230 GHz, 
they determined the background population of high-energy photons produced by inverse Compton scattering using the Monte-Carlo radiative transfer code \texttt{grmonty}. The pair production rate per unit volume is then determined from the distribution of seed photons using the Breit-Wheeler cross section $\sigma_{\gamma \gamma}$.
Recently, \citet{Wong2021} extended the work of \citet{Moscibrodzka2011} by modeling the pair drizzle scenario using radiation GRMHD simulations. In these simulations, \texttt{grmonty} is coupled to the GRMHD code and the photon distribution and number density is solved for self-consistently along with the plasma variables. 

The best-bet ``pair-drizzle" models in \citet{Moscibrodzka2011,Wong2021} predict maximal pair number densities in the funnel of $n_{\rm lep}=10^{-8}$ cm$^{-3}$ ($\approx 10^{-5}$ times the Goldreich-Julian density for Sgr A*) and $n_{\rm lep}=10$ cm$^{-3}$ (or $\approx 10^7$ times the Goldreich-Julian density for M87*). These quoted values for the lepton density $n_{\rm lep}$ are provided above the pole, where the simulation number density is determined by numerical floors. Hence, we cannot directly estimate a positron-to-electron ratio directly from these quoted values. Naively, we may assume the best estimate from these values is $f_{\rm pos}=1$, since the ion density is only non-zero in the polar region because of simulation floors, so the pair density from the \citet{Moscibrodzka2011,Wong2020} calculations can be seen as representing the only physical matter density along the poles. However, for the purposes of comparing to semi-analytic models, a more reasonable approach is to compare the calculated lepton density from the pair drizzle models to the electron number density in the 230 GHz emitting region. Neither \citet{Moscibrodzka2011} nor \citet{Wong2020} provide this information from their models, but we can estimate a reasonable order-of-magnitude number from the quoted accretion rates  \citet{Moscibrodzka2011}: $\dot{M} = 2\times 10^{-8} \dot{M}_{\rm Edd}$ for Sgr A* and  $\dot{M} = 10^{-6} \dot{M}_{\rm Edd}$ for M87*.\footnote{We define the Eddington accretion rate $\dot{M}_{\rm Edd}= L_{\rm Edd}/0.1c^2$, where the Eddington luminosity is $L_{\rm Edd} = 4\pi GMm_{\rm p}c/\sigma_{\rm T}$}. At a given radius $r$, assuming spherical symmetry, we can then estimate: 
\begin{equation}
    n_{\rm emit} \approx \frac{\dot{M}}{4\pi r^2 v^r m_p},
\end{equation}
where for the infall velocity we can assume $v^r = c/\sqrt{r/r_{\rm g}}$. We calculate the density at a radius $r=5r_{\rm g}$, which is a reasonable estimate of the emission radius for M87* \citep{Akiyama:2019fyp}. We then find from the quoted simulation accretion rates that $n_{\rm emit}\approx1.4\times10^3$ cm$^{-3}$ for Sgr A* and $n_{\rm emit}\approx2.2\times10^4$ cm$^{-3}$ for M87*. Using these simple estimates, we can compute positron-to-electron fractions of:
\begin{equation}
    f_{\rm pos} \approx 4\times 10^{-3} \;\;\; (\mathrm{M87^*}),
\end{equation}
and
\begin{equation}
    f_{\rm pos} \approx 10^{-13} \;\;\; (\mathrm{Sgr \, A^*}).
\end{equation}
These estimates are approximation. They depend sensitively on where the 230 GHz emission originates in the accretion flow and what the produced pair density and background electron density there are. For the pair density, we have chosen only one value along the pole from \citet{Moscibrodzka2011}. For the background electron density we have chosen an estimate motivated by spherical symmetry rather than a precise measurement from the emission location. Furthermore, note that both \citet{Moscibrodzka2011} and \citet{Wong2020} focus only on one type of accretion flow with weak magnetic flux on the black hole. The M87* accretion flow, in contrast, is expected to be strongly magnetized (in the MAD state \citealt{Akiyama:2021fyp}). Nonetheless, we take these estimates as motivating values for our choices of $f_{\rm pos}$ which we explore in the models below. 

\section{Semi-Analytic Models}
\label{sec:models}
Our primary objective in this work is to develop representative semi-analytic models for polarized emission from jet/accretion flow systems that incorporate the effects of a non-zero positron fraction. we are particularly  interested in predictions from these models for the polarized submillimeter emission structure that may be resolved by the EHT. The strategy we employ is to modify existing radiative transfer pipelines to allow for a non-zero positron content, along with other physical parameters such as the spatial distribution of particles, the non-thermal particle fraction, and the parameterized energy distribution function of the emitters.  

Here, we discuss the form of the models we use to model emission from M87* and Sgr A*. While in the next Section, we discuss the implementation of positron effects we use in our radiative transfer code. 

\subsection{M87*: Semi-Analytic Jet Model}
\label{subsec:JetFluidModel}
For M87*, we use a self-similar, semi-analytic jet model calibrated to measurements from a GRMHD simulation. Our semi-analytic jet model is based on that introduced in \cite{Anantua2020a}. However, we extend this model in significant ways: we add 
general relativistic ray tracing, we consider non-thermal distributions with arbitrary power-law indices $p\neq 2$, we map the particle number density to the total electron/positron pressure rather than the partial pressure of particles contributing within an octave of the observed frequency, and we consider overall higher electron-to-magnetic pressure ratios $\beta_{\rm e}$. Here we summarize the full model description including these changes.

The semi-analytic jet model assumes self-similarity in the jet parabolic parameter $\xi=s^2/z$, where $(s,z,\phi)$ are cylindrical coordinates. The model is determined by three fitting functions in $\xi$ we determine from GRMHD simulation data. These fitting functions are for the magnetic flux threading the jet cross-section $\Phi_B(\xi)$, the field line angular speed $\Omega_B(\xi)$, and the $z$-component of the velocity vector $v_z(\xi)$.
 
We fit to these quantities in the finite simulation region, then extrapolate to large radii through self-similarity in $\xi$. 

The {\tt HARM} simulation \citep{McKinney2012} we use as the basis of our semi-analytic model is a magnetically arrested disk (MAD) accreting on a black hole with spin $a/M=0.92$, and the fitting forms for plasma variables in the self-similar model \cite{Anantua2020a} are computed at altitude $z=50M$. The simulation outputs code units in terms of arbitrary scales for the black hole mass $M$ and field strength $B$. Physical quantities such as the length scale $r_{\rm g}\equiv GM/c^2$ and time scale $t_\mathrm{g}\equiv GM/c^3$ are then determined through specifying the black hole mass $M_\mathrm{BH}$ and the magnetic field scale $B_0$ such that we obtain the observed flux of 1.5 Jy for M87* \citep{Akiyama:2019bqs}. \footnote{An alternative procedure for code unit conversion for the magnetic field can be found in Table \ref{Tab:CodeToPhysicalUnits} in Appendix \ref{EstimatingJetBeta} of \cite{Anantua2020a}).}
We fix the M87* black hole mass $M_{\mathrm{87}} = 6.5\times 10^9 M_\odot$ \citep{Akiyama:2019bqs}, which gives $r_{\rm g} = 9.6\times10^{14}$cm and $t_{\rm g} \approx 9\mathrm{\ hr}$. 

The fitting forms for the magnetic flux, field line angular speed, and $z-$component of the velocity
are adapted from \citet{Anantua2020a} (given in dimensionless code units):
\begin{align}
 \Phi_B(\xi)& = \tanh(0.3\xi),\label{RRJetPhi}\\
 \Omega_B(\xi)&= 0.15e^{-0.3\xi^2},\label{RRJetOmega}\\
 v_z(\xi,z) &= \left[-1.2 + 2.2\mathrm{tanh}(0.85\log z)\right]e^{-0.001\xi^4}. \label{eq:RRJetvz}
\end{align} 
We set $\xi_\mathrm{max}=10$ as the boundary of the jet. Given these fitting forms, the magnetic field in cylindrical coordinates is:
\begin{align}
\label{eq:bfield}
B_s &= B_0 \frac{s\Phi_B'(\xi)}{2\pi^2}, \\
B_z &= B_0 \frac{\Phi_B'(\xi)}{\pi z}, \\
B_\phi &= B_0 \frac{-s\Omega_B(\xi)\Phi_B'(\xi)}{\pi z},
\end{align}
where $B_0$ is the scaling factor between dimensionless code units and physical units that we measure by the fixing the total flux density of the 230 GHz image.
Note that the total magnetic flux in the jet is conserved with height $z$. If we knew the total magnetic flux in the jet $\Phi_{H}$, we could instead use it to determine $B_0$ via:
\begin{equation}
\label{eq:flux}
    \Phi_{H}=\int_0^{s_\mathrm{max}(z)}2\pi s B_z ds \approx B_0 r_g^2.
\end{equation}
Once we have determined the magnetic field at a point in the jet and converted it to the cgs units, the magnetic pressure throughout the jet is read as:
\begin{equation}
P_B = \frac{B^2}{8\pi}.
\end{equation}
The three-velocity vector in this model is:
\begin{equation}
\begin{pmatrix} v_s  \\ v_\phi  \\ v_z
 \end{pmatrix}
=\begin{pmatrix} \frac{s}{2z}v_z(z,\xi)  \\ s\Omega_B\left(1-v_z(z,\xi)\right)  \\ v_z(z,\xi)
 \end{pmatrix},
\end{equation}
where the fitting function for $v_{z0}(z)$ is given by Eq~\ref{eq:RRJetvz}.

Considering the model before adding pairs, we assume the emitting electrons have a power-law distribution function between a minimum Lorentz factor $\gamma_{\rm min}$ and a maximum Lorentz factor $\gamma_{\rm max}$:
\begin{equation}
    N_{e^-}(\gamma) = 
\begin{cases}
N_0\gamma^{-p} &\gamma_{\rm min}\leq \gamma\leq\gamma_{\rm max}\\
0 & \mathrm{otherwise}
\end{cases},
\label{eq:powerlaw}
\end{equation}
where $N_0 = n_{e^-} (p-1)/\left(\gamma_{\rm min}^{1-p}-\gamma_{\rm max}^{1-p}\right)$ is the overall normalization for the distribution and $n_{e^-}$ is the total electron number density. 
In our emission model, we determine the electron pressure throughout the jet by specifying the electron pressure to magnetic pressure ratio $\beta_{\mathrm{e}} = P_{e^-}/P_B$ throughout the emission region. In this work, we assume that the $\beta_e=\beta_{e0}$ is constant. The total electron number density for the power-law distribution in Equation~\ref{eq:powerlaw} is then give by:
\begin{equation}
\label{eq:constbeta}
   n_{e^-} = \frac{3}{m_ec^2}\frac{(p-2)}{(p-1)}\frac{\gamma_{\rm min}^{1-p} - \gamma_{\rm max}^{1-p}}{\gamma_{\rm min}^{2-p} - \gamma_{\rm max}^{2-p}}\beta_e \, P_B.
\end{equation}
In our emission model, we propose that each emitting positron has the same distribution function in the $\gamma$ space, but scaled according to a different total number density. The ratio of positrons to electrons: \begin{equation}
    f_\mathrm{pos}\equiv \frac{n_{e^+}}{n_{e^-}}, 
\end{equation}
is taken to be constant throughout the emitting region and is a free parameter in our models.
The total positron number density $n_{e^+}$ is then given by $n_{e^+}=f_{\mathrm{pos}}%f_{e^+}
n_{e^-}$.

\subsection{Sgr A*: Radiatively Inefficient Accretion Flow Model} \label{subsec:DiskFluidModel}
For modelling Sgr A* images and spectra, we adapt the radiatively inefficient accretion flow (RIAF) model first presented 
in\cite{Broderick2006a,Broderick2006b,Broderick2009,Broderick2016}. 
This choice is motivated mainly from the most recent observations of Sgr A* from \cite{2019ApJ...871...30I}, who showed that disk dominated models of Sgr A* are better matched with the observations than jet models. \citet{2019ApJ...871...30I} found that jet dominated models are severely constrained at 86 GHz, and the only jet models in their tested simulations that are consistent with the measured 1.3 mm and 3 mm image sizes and asymmetry are viewed nearly face-on at $\lesssim20^\circ$. In a future work we will consider hybrid disk-jet models for both Sgr A* and M87*.

Before considering pairs, the \citet{Broderick2006a} RIAF disk model features axisymmetric, power-law spatial distributions of both thermal electrons (with number density $n_{\mathrm{th}}$ and temperature $T_e$) and nonthermal electrons (with number density $n_{\mathrm{nth}}$, power-law index $p$, minimum Lorentz factor $\gamma_{\rm min}$, and maximum Lorentz factor $\gamma_{\rm max}$). Following \citet{Broderick2009}, we define the electron temperature, thermal and nonthermal number densities as power-laws:
\begin{equation}
\begin{tabular}{cr}
  $n_\mathrm{th} = n_{0,\mathrm{th}}\left(\rho/2\right)^{-1.1}e^{\left(-z^2/2s^2 \right)} 
  $ ,
  \\$   T_e = T_0 \left(\rho/2\right)^{-0.84}$ ,
  \\
  $n_\mathrm{nth} = n_{0,\mathrm{nth}}\left(\rho/2\right)^{-2.02}e^{\left(-z^2/2s^2\right)}$,
  \end{tabular}\label{Eq:DiskFluidModel}
\end{equation}  
where $\rho$ is the spherical radius while $s = \sqrt{\rho^2-z^2}$ is the cylindrical radius. We take both $T_0$ and the nonthermal electron fraction $f_{\rm nth} = n_{0,\mathrm{nth}}/n_{0,\mathrm{th}}$ as free
parameters in our model. We fix $n_{0,\mathrm{th}}$ by requiring that our model produces the observed 3.5 Jy 230 GHz flux density for Sgr A* \citep{Bower2015}.

The magnetic field strength in the \citet{Broderick2006a} model is toroidal everywhere. The overall field strength is set by a total plasma $\beta$, which we take as a free parameter: 
\begin{equation}
\frac{B^2}{8\pi}=\frac{1}{\beta}\frac{n_\mathrm{th}m_pc^2}{24 r}.
\end{equation} 
The fluid velocity is Keplerian outside of the innermost stable circular orbit (ISCO) radius. While the fluid is in free-fall inside the ISCO.

Exactly as in the M87 jet model, if a positron population is present, we assume it has the same properties as the thermal+nonthermal electron distribution with a different overall density normalization. The ratio of positron-to-electron number densities  $f_\mathrm{pos}=n_{e^+}/n_{e^-}$ is again fixed throughout the plasma and is taken as a free parameter in our models. 

\section{Implementation and model space}
\label{sec:modelspace}
In this Section, we describe the implementation of our semi-analytic jet and RIAF models in the public \texttt{GRRT} code \texttt{GRTRANS} and we summarized our strategy for surveying the model space available to each model. 

\subsection{Radiative Transfer with \texttt{GRTRANS}}
Polarized radiative transfer through an astrophysical plasma must account for the effects of differential polarized emissivities, absorption, Faraday conversion of linear to circular polarization, and Faraday rotation of the linear polarization EVPA. In a curved spacetime, the emitted light received by the observer is also modified by the curvature of photon trajectories and the parallel transport of the polarization basis along geodesics. 

Several numerical codes exist that perform polarized radiative transport in the Kerr spacetime. These include e.g. \texttt{IPOLE} \citep{Moscibrodzka2017}, \texttt{BHOSS} \citep{younsi2020}, and \texttt{GRTRANS} \citep{Dexter2016}, which we use in this work. \texttt{GRTRANS} first solves for the geodesic trajectories between the pixels on the observer screen and the emitting region around the black hole. It then performs radiative transfer along these geodesics to solve for the Stokes parameters $I, Q, U, V$. It assumes that the radiative transfer on different geodesics is independent, and so cannot account for scattering processes. In this work, we consider synchrotron radiation only. 

In the local, emitting frame, the full polarized radiative transfer equations for the Stokes parameters $I, Q, U, V$ take the form:
\begin{align}\label{PolarizedRad}
\frac{d}{d s} \begin{pmatrix} I'  \\ Q'  \\ U' \\ V'
  \end{pmatrix}  = \begin{pmatrix} j'_I  \\ j'_Q  \\ j'_U \\ j'_V\end{pmatrix}  - \begin{pmatrix} \chi'_I & \chi'_Q  & \chi'_U &  \chi'_V \\
 \chi'_Q &  \chi'_I  &  \rho'_V  & \rho'_U \\
   \chi'_U  & -\rho'_V &  \chi'_I  & \rho'_Q   \\
     \chi'_V  & -\rho'_U & -\rho'_Q  & \chi'_I 
\end{pmatrix} \begin{pmatrix} I'  \\ Q'  \\ U' \\ V'
  \end{pmatrix},
\end{align}
where the $j$s refer to emissivities, the $\chi$ terms are absorption coefficients, and the $\rho$ describe the "rotativities" that mix linear and circular polarizations. For instance, $\rho_V$ is the Faraday rotation coefficient that rotates the plane of linear polarization, while $\rho_Q$ is the Faraday conversion coefficient that converts linear into the circular polarization. \texttt{GRTRANS} simplifies these equations by rotating the emitting frame so that $j_U, \chi_U, \rho_U=0$. \texttt{GRTRANS} also accounts for Doppler boosting of the emitted intensities from the radiation to the lab frame, as well as the parallel transport of the polarization frame (which can be specified in terms of a rotation of the Stokes basis in Equation~\ref{PolarizedRad}, see \citet{Dexter2016} Section 2.2). 

The $j,\chi$ and $\rho$ coefficients we use are all for synchrotron radiation. They depend on the electron distribution function $N(\gamma)$ as well as the magnetic field strength and orientation relative to the wavevector in the fluid rest frame. In this work, we consider both thermal (for the RIAF model) and non-thermal (for both the jet and RIAF models) lepton distributions. The coefficients we use in \texttt{GRTRANS} are defined in Appendix of \citet{Dexter2016} for both the thermal and nonthermal distributions. 

In all of our models, we assume a constant positron-to-electron number density ratio $f_\mathrm{pos}$. To include the effects of positrons in the radiative transfer, we simply modify the emissivity and rotativity coefficients as follows: 
\begin{equation}
\begin{tabular}{cr}
  $j_{I,Q,U}\to\left(1+f_{e^+}\right)j_{I,Q,U}$,
  \\$j_V\to \left(1-f_{e^+}\right) j_V$ ,
   \\
  $\chi_{I,Q,U}\to\left(1+f_{e^+}\right)\chi_{I,Q,U}$,
    \\$\chi_V\to \left(1-f_{e^+}\right) \chi_V$ ,
       \\
  $\rho_{Q,U}\to\left(1+f_{e^+}\right)\rho_{Q,U}$,
    \\$\rho_V\to \left(1-f_{e^+}\right) \rho_V$.
 \end{tabular}
 \label{eq:modemis}
\end{equation}  
Equation~\ref{eq:modemis} indicates that e.g. the emissivities for total intensity and linear polarization increase in magnitude with the addition of emitting positrons, as their total intensity and linear polarization emission is indistinguishable from that of electrons. The circular polarization emissivity is decreased by the addition of positrons as positrons emit circular polarization in the opposite sense to electrons. However, the Faraday conversion coefficient which determines how linear polarization is converted into the circular polarization, $\rho_Q$, is \emph{increased} in magnitude (with the same sign) in the presence of positrons, while the Faraday rotation coefficient $\rho_V$ \emph{decreases}; there is no Faraday rotation in a pure pair plasma. 

\subsection{M87* Jet Model Parameters}
\label{sec:constbeta}
In producing image of our jet model for M87*, we assume a dimensionless black hole spin of $a/M=0.5$ and a viewing angle of $i=20\deg$ \citet{Akiyama:2019fyp}. We produce resolved images at 230 GHz and 86 GHz, as well as spectra from $10^9-10^{16}$ Hz.

In producing model images and spectra from the M87* jet model, we fix the electron-to-magnetic pressure ratio $\beta_e$ to be constant throughout the jet (Eq.~\ref{eq:constbeta}):
\begin{equation}
    \beta_e=\beta_{e0}.
\end{equation}
This simple emission prescription assumes the relativistic electron pressure (which we define by \emph{not} including the positron contribution) is a constant fraction $\beta_{\mathrm{e0}}$ of the local magnetic pressure everywhere. We call this model the ``Constant Electron Beta Model," following \cite{Blandford:2017pet,Anantua2018}.\footnote{ 
Note that unlike \cite{Anantua2020a}, we drop prefactors $\frac{n-n_i}{n}=\frac{n_{e^-}+n_{e^+}}{n}$ to simplify the changes in the \texttt{GRTRANS} code when incorporating positrons.} 

We survey a space of constant electron beta values in the range: 
\begin{equation}
\beta_{\mathrm{e0}} \in \{ 10^{-6},10^{-4},10^{-2}\} 
\end{equation}  
We also vary the power-law index $p$ of the emitting distribution: 
\begin{equation}
    p \in\{2.5,2.8, 3.0,3.2, 3.5\}.
\end{equation}
Finally, we also vary the positron fraction in the range:
\begin{equation}
    f_\mathrm{pos} \in\{0.0, 0.01, 0.1, 0.5, 1.0 \}.
\end{equation} 
This range of positron values spans the range from pure ionic to pure pair plasmas and allows us to characterize the effect of adding pairs across a large range of parameter space. However, we focus on small values of the positron fraction $\lesssim 10\%$, motivated by theoretical estimates for the positron fraction in M87* and Sgr A* from various pair production mechanisms (Section~\ref{sec:estimates}), \citep{Moscibrodzka2011,Broderick2015,Wong2021}.
In \cite{Anantua2020b}, we found that the constant $\beta_{e0}$ model could produce low circular polarization fractions $V/I$ of $\mathcal{O}(10^{-3})$ with low $\beta_{\mathrm{e0}}=10^{-10},10^{-8},10^{-6}$.
For the higher $\beta_{\mathrm{e0}}$ values which we use in the current work, we see a different interplay of coupled emissivities, absorption coefficients and rotativities-- resulting in significant Faraday effects and more circular polarization.

Throughout this work, we vary $\beta_{\mathrm{e0}}$, $p$, and $f_{\mathrm{pos}}$. We fix all other parameters, such as the minimum and maximum Lorentz factors: $\gamma_{\mathrm{min}} = 10$, and  $\gamma_{\mathrm{max}} = 5 \times 10^3$. We determine the magnetic field scale $B_0$ by finding the value that produces the observed total flux from M87* at 230 GHz. 

\subsubsection{Sgr A* RIAF model parameters}
For modeling Sgr A*, we use the RIAF model from \cite{Broderick2006a} described in Eqs. \ref{Eq:DiskFluidModel}. We fix the fiducial dimensionless black hole spin as $a= 0.1$, inclined $60^\circ$ relative to the line of sight, motivated by the proto-EHT model fitting results from \citet{Broderick2016}.

In our implementation, the \citet{Broderick2006a} RIAF model is determined by seven plasma parameters: the normalization of the thermal and nonthermal number density profiles $n_{0,\mathrm{th}}$ and $n_{0,\mathrm{nth}}$, the normalization of the temperature profile $T_0$, the plasma $\beta$ that sets the $B$-field strength, the power-law slope of the nonthermal distribution $p$, and the minimum and maximum nonthermal Lorentz factors $\gamma_\mathrm{min}$ and $\gamma_\mathrm{max}$.

Because current EHT polarimetric constraints are stronger for M87* than for Sgr A*, we explore a smaller parameter space in the Sgr A* RIAF model. Our goal is not to find a best-fit RIAF model, but to explore the trends in the model across a few parameters of interest when adding positrons to the emitting plasma. In particular, we fix $T^0_e = 1.5 \times 10^{11} K$. We determine the non-thermal part of the electron number density through the ratio $f_{\rm nth}$ by $n_{0,\rm{nth}} = n^0_e f_{\rm{nth}}$, and we fix the value $f_{\rm{nth}}=0.01$. We also fix $p=2.8$ and $\gamma_{\rm min}=100$ and $\gamma_{\rm max}=10^6$. Once all other model parameters are set, we fix the thermal number density $n_{0,\mathrm{th}} $ by requiring the total flux at 230 GHz $F_{230} = 3.5 \rm{Jy}$. 

The two parameters we explore in the RIAF model are $\beta$ and $f_{\rm pos}$. We explore plasma-beta in the range:
\begin{equation}
\beta\in\{0.01, 0.1, 1.0, 10.0\},
\end{equation}
to explore sub-equipartition, equipartition and super-equipartition plasmas. The positron fractions we explore span the same range as the M87* jet model; which is:
\begin{equation}
    f_\mathrm{pos} \in\{0.0, 0.01, 0.1, 0.5, 1.0 \}.
\end{equation}

\section{Model Exploration and Constraints}
\label{sec:Model-search}
After setting up semi-analytic models for M87* and Sgr A* that include positrons in the emitting lepton population in Section~\ref{sec:models} and describing the model space we explore in Section~\ref{sec:modelspace}, we scan the parameter space for each model defined in Section~\ref{sec:modelspace}, exploring the physics of different emitting particle distributions, relative magnetic field strengths, and positron content. 
For each model parameter set, we produce 230 GHz Stokes maps and polarized spectra from radio to X-ray frequencies.

In this section, we discuss constraints from observations of M87* and Sgr A* that we apply on these results from each model. In particular, for both M87* and Sgr A*
we compute reduced-$\chi^2$ goodness-of-fit statistics to the overall SED 
and disfavor models with 
the largest $\chi^2$s. We also compare 230 GHz images of our M87* jet model to the most recent polarization constraints from  \citet{2021ApJ...910L..13E} and construct a pass/fail table of models according to whether they satisfy the EHT constraints 
Finally, we determine best-bet models which best satisfy both sets of observational constraints. For Sgr A*, on the other hand, we limit our study to a smaller number of models similar to the ones fit to the proto-EHT data in \citet{Broderick2016}. We add the effects of positrons to the emission model in this setup and discuss which models are closest to the observational constraints without determining a ``best-bet" fiducial model for the polarized emission.

\subsection{Total Intensity SED $\chi^2$ values}
As a first test of both the M87* constant beta model and Sgr A* RIAF model, we construct $\chi^2$ factors comparing the predicted SED from each model to the observational data from each source.  We use total intensity SED data from radio to X-ray frequencies, as described in Section~\ref{sec:observations}. For each model, we use this test to find a reasonable set of preferred models from the larger parameter space we explore. 

For the M87* jet model, we study a family of 75 models varying $\left( \beta_{\mathrm{e0}}, f_{\mathrm{pos}}, p \right)$. 
Our analysis of the total intensity SEDs from these models indicate that models with electron spectral index $p=3.5$ best fit the near-infrared SED data, as we expect from estimating the power-law index directly from the data in a one-zone power-law synchrotron emission model. Furthermore, the fit to the total intensity SED gets progressively better when we increase the lepton pressure by increasing $\beta_{\mathrm{e0}}$. Table 
\ref{chi2-constant-beta} presents the $\chi^2$ values for total intensity SED for all of the M87* jet models we consider. Because the observational SED is sparse in the interval $10^{12}\mathrm{Hz }<\nu<10^{13}\mathrm{Hz }$, we report the $\chi^2$ values for the radio and near-infrared parts of the spectrum separately, along with the overall $\chi^2$ from all data points. Recall that we set the magnetic field scale $B_0$ by fitting the model to the 230 GHz total flux of 1.5 Jy for M87*. 

We also study the total intensity SED $\chi^2$ for the Sgr A* RIAF model. Here we fix most parameters from  \cite{Broderick2016} and only vary two parameters: 
$\beta, f_{\mathrm{pos}}$.  The results of the Sgr A* SED fit analysis are given in Table \ref{RIAF-Model}. Recall that for each model we set the thermal electron number density $n_{0,\mathrm{th}}$ by fitting the model to the 230 GHz total flux of 3.5 Jy for Sgr A*. 

We emphasize that in this work, our primary aim is not to determine a single best fit model in our model space for either Sgr A* or M87* or to claim that either of the models we put forward for these two sources are the most complete or physically representative. Our aim is to find a ``reasonable" model for each source, chosen from a well-motivated set of parameters, and to explore the observational effects of changing the positron content in these models. In a future work, we will analyze the full parameter space available to each model more completely and determine a set of best fit parameters and their uncertainties with an MCMC posterior exploration code. 

\subsection{230 GHz Polarimetric Constraints}
From Table \ref{chi2-constant-beta}, it is evident that the M87* jet model $\chi^2$ constructed from the total intensity SED does not show a strong dependence on the positron fraction.
To place constraints on $f_{\rm pos}$ in these models, we need to turn to polarimetric data constraints. M87* and Sgr A*'s polarized SED's are much more poorly constrained than the SED in total intensity. Here, we only use the most recent observational constraints for the linear and circular polarization at 230 GHz to distinguish between models. We find, similarly to the M87* GRMHD analysis in \citet{Akiyama:2021fyp}, that these polarimetric constraints break degeneracies between different models that equally satisfy the total intensity observations.

\subsubsection{M87* 230 GHz polarimetric constraints}
For M87*, we use the most recent polarimetric
constraints at 230 GHz from \cite{Akiyama:2021fyp}. While 
that work reports the resolved and unresolved fractional linear polarization, it only provides an upper limit on the circular polarization from simultaneous ALMA observations during the 2017 EHT campaign \citep{Goddi2021}. In summary, the most recent observational constraints on the fractional linear and circular polarizations defined in Eqs. ~(\ref{eq:polfracs1}-\ref{eq:polfracs2}): 
\begin{align}
\label{I-Q-U-V}
1\%  \leq |m|_{\mathrm{net}} \leq  3.7 \%,    \nonumber\\
5.7 \%  \leq  \langle |m| \rangle  \leq 10.7 \%,  \nonumber\\
0\%  \leq |v|_{\mathrm{net}}   \leq 0.8 \% ~. 
\end{align} 
The EHT ranges for the polarimetric ratios above are conservative, incorporating the results from several image reconstruction techniques on the M87* data. However, these ranges do not incorporate source variability on timescales longer than $\sim1$ week, which may be significant. 
As a consequence, we score models against the EHT constraints in two ways. First, we use the direct EHT ranges from \citet{Akiyama:2021eap} as strict constraints -- none of our models pass these constraints directly. Second, we interpret the EHT ranges as a central value and $1\sigma$ error region and consider whether any models pass within $1.5\sigma$ (equivalent to expanding the allowed range by 50\% from the mid-point to account for possible future source variability). Here we find some of our models do satisfy this looser constraint.  
In Table \ref{P-V-constant-beta} in 
Appendix \ref{P-V-Observation}, we 
compute the aforementioned quantities in our different models and we give scores of pass (P) or fail (F) to individual models based on their performance on all three constraints. Notably, only 4\% of models pass all of the constraints (Section \ref{fiducial-constant-beta}) at 1.5$\sigma$.

\subsubsection{Sgr A* 230 GHz polarimetric constraints}
Our exploration of the Sgr A* RIAF model space is more limited than for the M87* jet model, and we mostly use this model to illustrate physical effects of adding positrons of a disk-dominate system. However, we do make a preliminary comparison of Sgr A* to polarimetric data, using observational constraints from \cite{2015Sci...350.1242J, 2018ApJ...868..101B}. These constraints are: 
\begin{align}
\label{I-Q-U-V}
|m|_{\mathrm{net}}   \simeq & 7\%,    \nonumber\\
10 \%  \leq  \langle |m| & \rangle  \leq   30 \%,  \nonumber\\
|v|_{\mathrm{net}} \leq & 1.0 \% ~. 
\end{align}
Since our parameter space is more limited and these data constraints will soon be supplemented with the results from the EHT 2017 observations, 
in what follows, we do not 
attempt to make a pass-fail table for our sample of RIAF models, and we leave a detailed evaluation of the RIAF models to future studies. Instead, we take a more qualitative approach and discuss the implications of these initial constraints on the RIAF models. We provide values of $|m|_{\rm net}$, $|v|_{\rm net}$ and $\langle|m|\rangle$ for the Sgr A* RIAF models in Table \ref{P-V-RIAF}.

\subsection{Best-Bet Models}
Having applied total intensity SED and 230 GHz polarimetric constraints to the models, here we determine the passing ``best-bet" models for the M87* jet.
For the Sgr A* RIAF model, on the contrary, we choose the family of models that are closer to the current constraints, while we leave a more detailed analysis for future work. 

\subsubsection{M87* jet model}
\label{fiducial-constant-beta}
We choose our best-bet model from the initial set of 75 models for M87* by selecting models that have a low total $\chi^2$ when comparing to the total intensity SED and that also satisfy the current 230 GHz polarimetric constraints as described above. In general, models with higher $\beta_{\mathrm{e0}}$ do a better job fitting the total intensity SED, and models with small values of $f_\mathrm{pos}$ are generally closer to the EHT polarimetric constraints. We settle on three fiducial best-bet models that we label A,B,C as follows: 
\begin{align}
\mathrm{Model} =& [f_\mathrm{pos}, \beta_{\mathrm{e0}}, p], \nonumber\\
A \equiv&  [0.0, 10^{-2}, 3.5], \nonumber\\
B \equiv & [0.01, 10^{-2}, 3.5], \nonumber\\
C \equiv &  [0.1, 10^{-2}, 3.5].
\end{align}
These models differ only in their positron fraction. From Table \ref{P-V-constant-beta}, it is evident that some of these only barely satisfy the above polarimetric constraints, but all perform significantly better than other models in the set, especially when compared with those with very low values of $\beta_{\mathrm{e0}} \sim 10^{-6}$ or with $f_\mathrm{pos} \gtrsim 0.5$.  The polarimetric constraints are very powerful in breaking the degeneracies between different models that have similar $\chi^2$ values in total intensity. 

\subsubsection{Sgr A* RIAF model}
In considering the Sgr A* RIAF model, we explore a smaller parameter space focused primarily on varying plasma-$\beta$ and the positron fraction $f_{\rm pos}$. We fix all other parameters to those fit in \cite{Broderick2016}. Our aim is not to completely explore the parameter space for this model, but to identify trends in the resulting images and spectra with different values of plasma $\beta$ and positron fraction $f_{\rm pos}$.

At the level of the total intensity SED $\chi^2$, models with different values of plasma-$\beta$ and positron fraction $f_{\rm pos}$ all fit the data relatively well. However, their predictions for the 230 GHz linear and circular polarizations are quite different. Applying the above constraints to our very limited family of RIAF models, we find that models with lower values of $\beta$ do a better job. Consequently, we pick the model with $\beta = 10^{-2}$ as our fiducial model. In the next section, we discuss the effects of varying $f_{\rm pos}$ in this model.

%%%%%%%%%%%%%%%%%%%%%%%%%%%%%%%%%%%%%%%%%%%%%%%%%%%%
\begin{figure*}[th!]
\center
\includegraphics[width=1.0\textwidth]{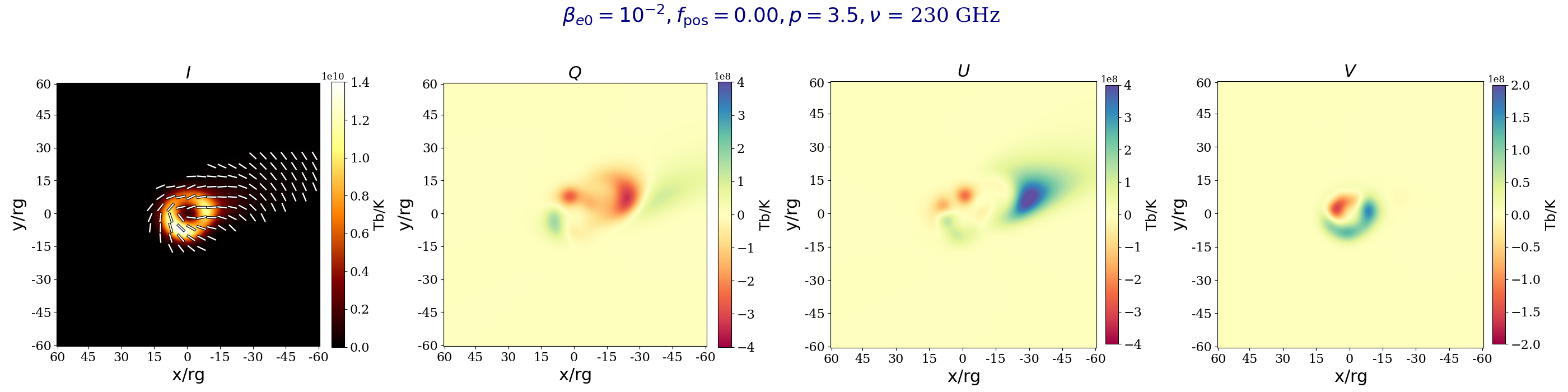}
\includegraphics[width=1.0\textwidth]{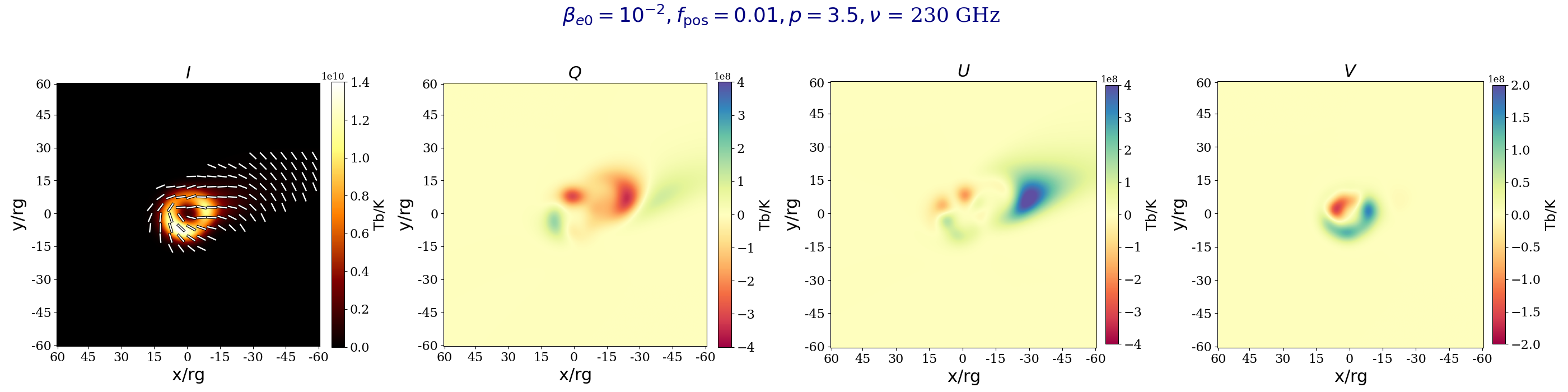}
\includegraphics[width=1.0\textwidth]{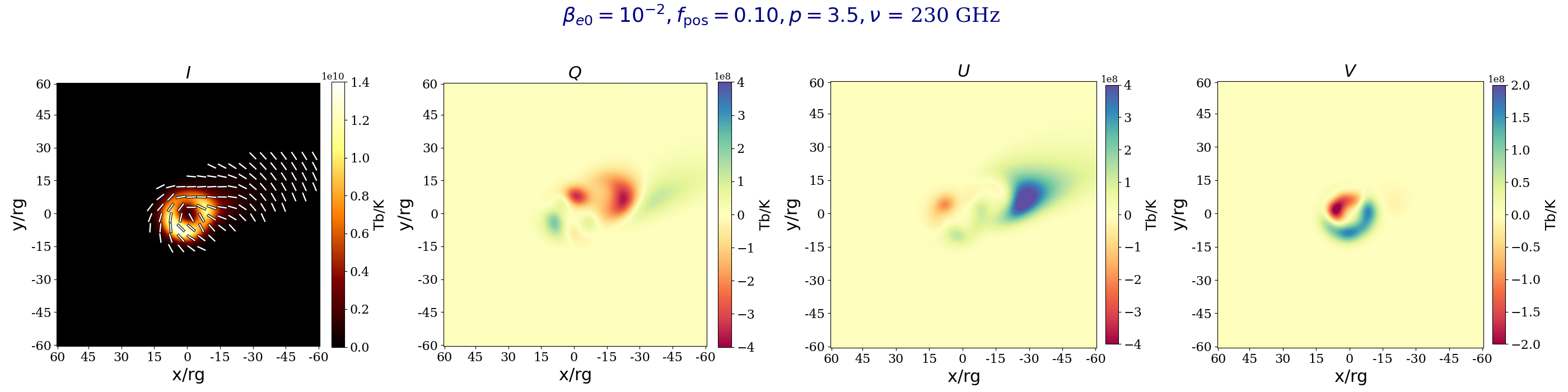}
\caption{ Stokes maps for the inferred best-bet M87* jet models in Sec. \ref{fiducial-constant-beta}. From the Top to Bottom, we increase the $f_{\mathrm{pos}}$ associated with models A-C, respectively. Before display, the images were blurred to the nominal EHT resolution by convolution with a circular Gaussian kernel with FWHM of 20 $\mu$as. The brightness temperature in each model is displayed in a linear scale. Ticks overlaid on the Stokes $I$ image denote the EVPA, and are plotted where the fractional linear polarization is above 0.5 \% of its maximum value. } \label{StokesMapsBetaE010Em2Pnth3Pt2Theta20fPos0AndPt5And1}
\end{figure*}
%%%%%%%%%%%%%%%%%%%%%%%%%%%%%%%%%%%%%%%%%%%%%%%%%%%%

%%%%%%%%%%%%%%%%%%%%%%%%%%%%%%%%%%%%%%%%%%%%%%%%%%%%
\begin{figure*}[th!]
\center
\includegraphics[width=1.0\textwidth]{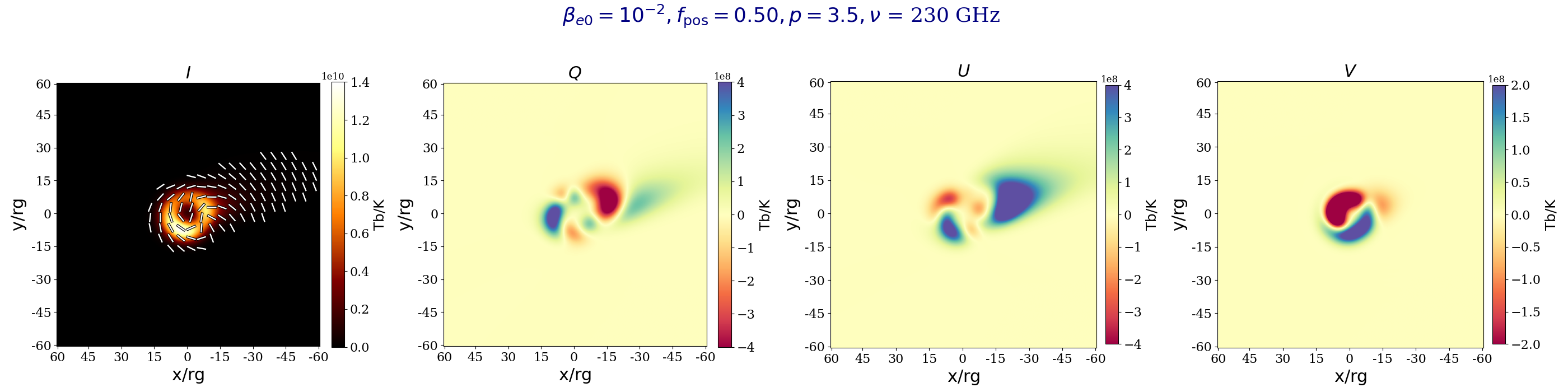}
\includegraphics[width=1.0\textwidth]{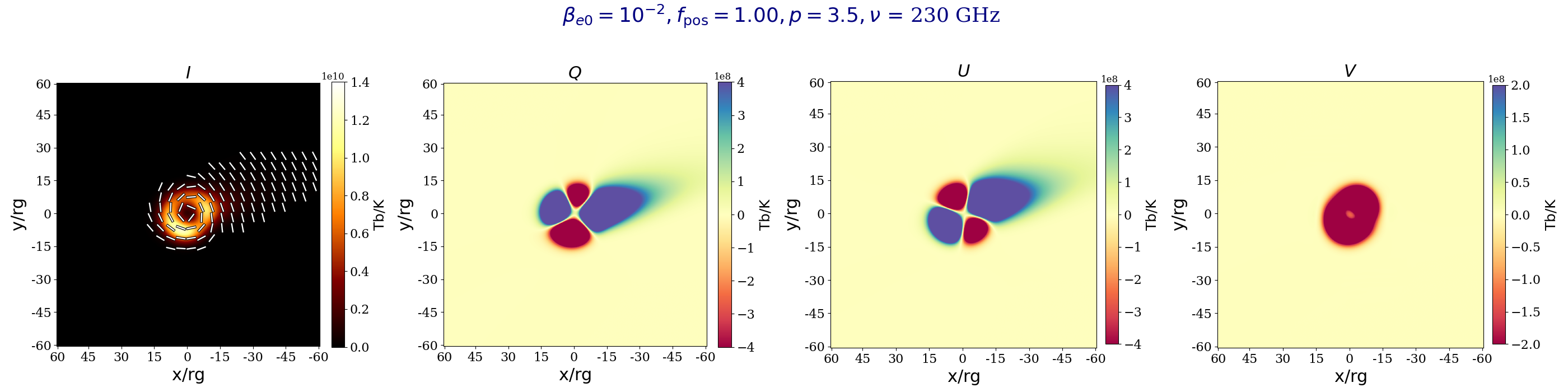}
\caption{ Stokes maps for M87* jet model with $f_{\mathrm{pos}} = (0.5, 1.0).$ The rest of the model parameters are fixed following the best-bet model. We plot all panels in the same style as in Fig.~\ref{StokesMapsBetaE010Em2Pnth3Pt2Theta20fPos0AndPt5And1} to facilitate comparison; as a result, the polarization maps for $f_{\rm pos}=1$ saturate the color scale.} \label{StokesMapBetaFullPos}
\end{figure*}
%%%%%%%%%%%%%%%%%%%%%%%%%%%%%%%%%%%%%%%%%%%%%%%%%%%%
\section{Results: Images and Spectra}\label{sec:Results}
Having chosen fiducial, ``best-bet" model parameters for M87* and Sgr A*,  here we discuss the polarized images and spectra from these models and their behavior when varying quantities of the interest, particularly the positron fraction $f_{\rm pos}$. For each family of models, we show several important examples in the main text and we present a more complete overview of the model space in Appendix~\ref{Beta-model-parameter-Exp}. 

\subsection{M87* constant $\beta_{\mathrm{e}}$ jet model}
We start with the constant $\beta_{\mathrm{e}}$ jet model for M87* and present 230 GHz images and polarized spectra for the best-bet models (A-C) defined in Sec.~\ref{fiducial-constant-beta}. Since the best-bet models vary only at the level of the positron fraction, we also explore the impact of pushing the positron fraction $f_{\mathrm{pos}}$ to large values out of the favored range from the EHT data constraints.

\subsubsection{Impact of $f_{\mathrm{pos}}$ on 230 GHz images}
\label{subsec:ConstBetaEModelMaps}
We first consider Stokes $I,Q,U,V$ maps of the 230 GHz emission from our favored constant $\beta_{\rm e0}$  jet models for M87*. Figure \ref{StokesMapsBetaE010Em2Pnth3Pt2Theta20fPos0AndPt5And1} presents Stokes maps for the fiducial models. Since in each case the magnetic field scale $B_0$ is determined such that the total flux matches observations at 230 GHz, we see few overall changes in the Stokes $I$ image structure and intensity. In each row, the images are blurred to the nominal EHT resolution by a convolution with a circular Gaussian kernel with FWHM of 20 $\mu$as.

From the plot, it is apparent that models A and B (with $f_{\mathrm{pos}} = 0.0, 0.01$) are visually difficult to distinguish, even in polarized emission maps. Moving to model C (with  $f_{\mathrm{pos}} = 0.1$), it is evident that the polarized intensities $Q, U$ and $V$ increase in magnitude relative to $I$. In all cases, our best-bet models feature a bilaterally asymmetric jet emanating from a ring around the central black hole, with polarization tracing the high intensity regions. The EVPA pattern around the ring is predominantly azimuthally spiraling as is observed in \citet{Akiyama:2021eap}. The fractional polarization of the jet emission is higher than the fractional polarization of the bright ring.

In Fig. \ref{StokesMapBetaFullPos}, we present the case with $f_{\mathrm{pos}} = 1.0$. Comparing Fig. \ref{StokesMapsBetaE010Em2Pnth3Pt2Theta20fPos0AndPt5And1} with Fig. \ref{StokesMapBetaFullPos}, we infer the following features about the impact of positrons on 230 GHz emission maps of M87:

$\bullet$ Increasing the positron fraction $f_\mathrm{pos}$ reduces the level of the bilateral asymmetry in the polarization maps as Faraday rotation diminishes. Stokes $Q$ and $U$ increase with increasing positron fraction as Faraday rotation is suppressed, the polarization pattern becomes more coherent, and beam depolarization is suppressed \citep[e.g.][]{Jimenez2018}.

$\bullet$ Faraday conversion is the dominant source of circular polarization in these models. Even in the pair plasma system the value of Stokes $V$ is nonzero and the Stokes $V$ emission map is qualitatively similar to the $f_{\rm pos}=0$ case, since the circular polarization is predominantly sourced by conversion. The linear polarization fraction is thus a better probe of the positron fraction in this model than the circular polarization fraction.

To justify our interpretation of the impact of Faraday effects in producing the bilateral asymmetry polarization and sourcing the Stokes V, in Appendix \ref{Faradays}, we show several examples in which we have manually turned off Faraday rotation and conversion effects for comparison with these results.
%%%%%%%%%%%%%%%%%%%%%%%%%%%%%%%%%%%%%%%%%%%%%%%%%%%%
\begin{figure*}
\center
\includegraphics[width=1.0\textwidth]{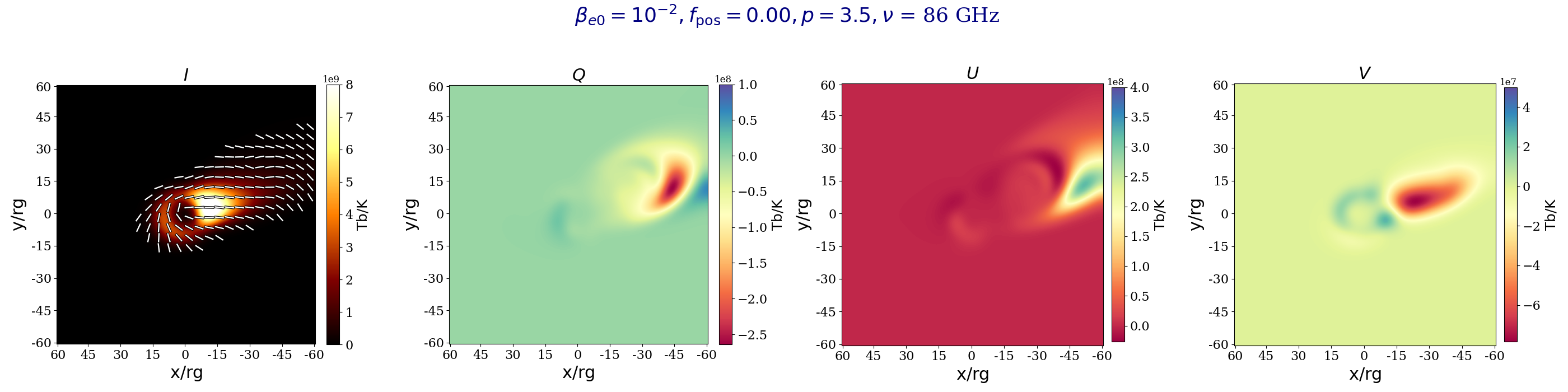}
\includegraphics[width=1.0\textwidth]{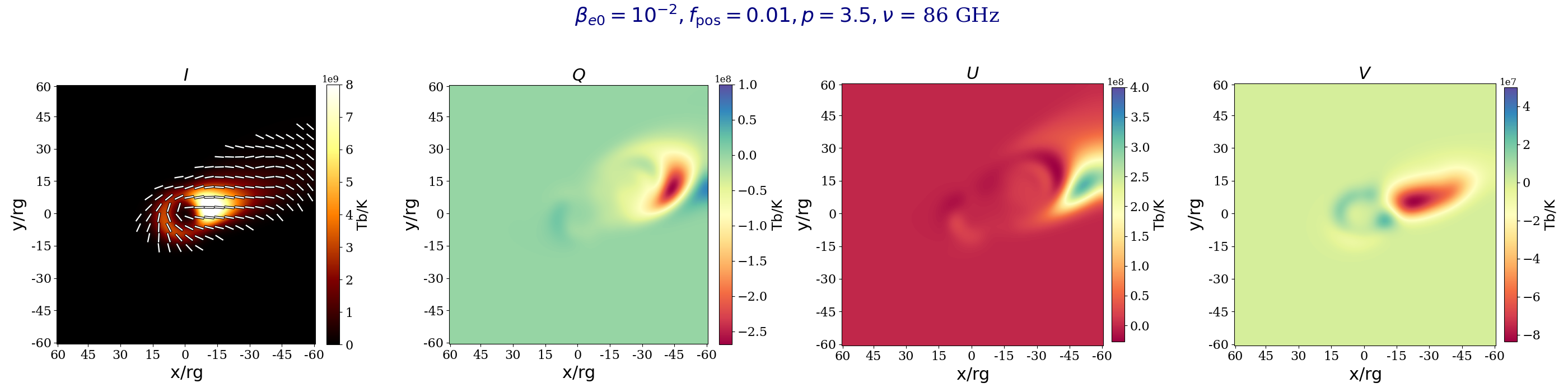}
\includegraphics[width=1.0\textwidth]{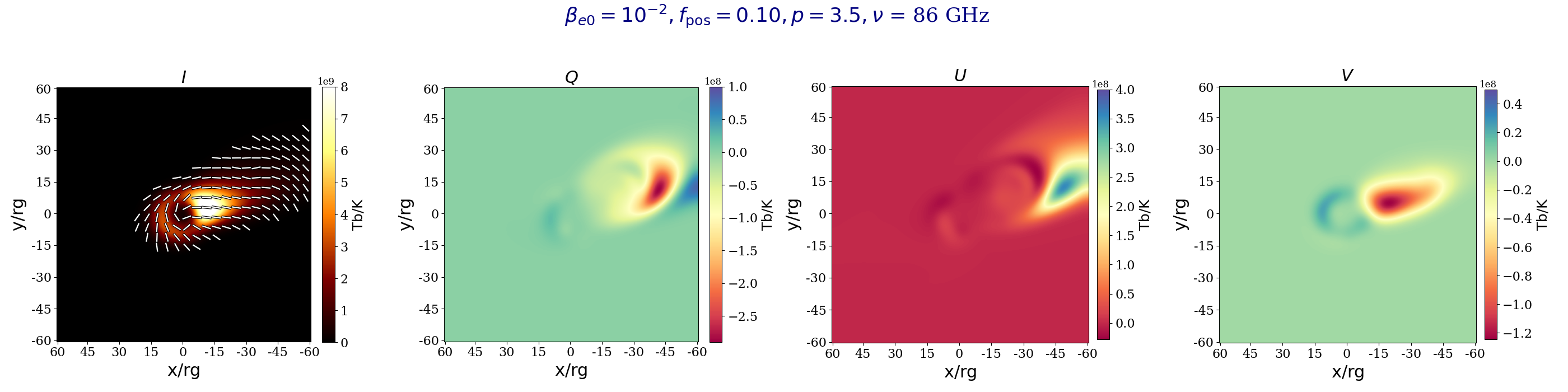}
\includegraphics[width=1.0\textwidth]{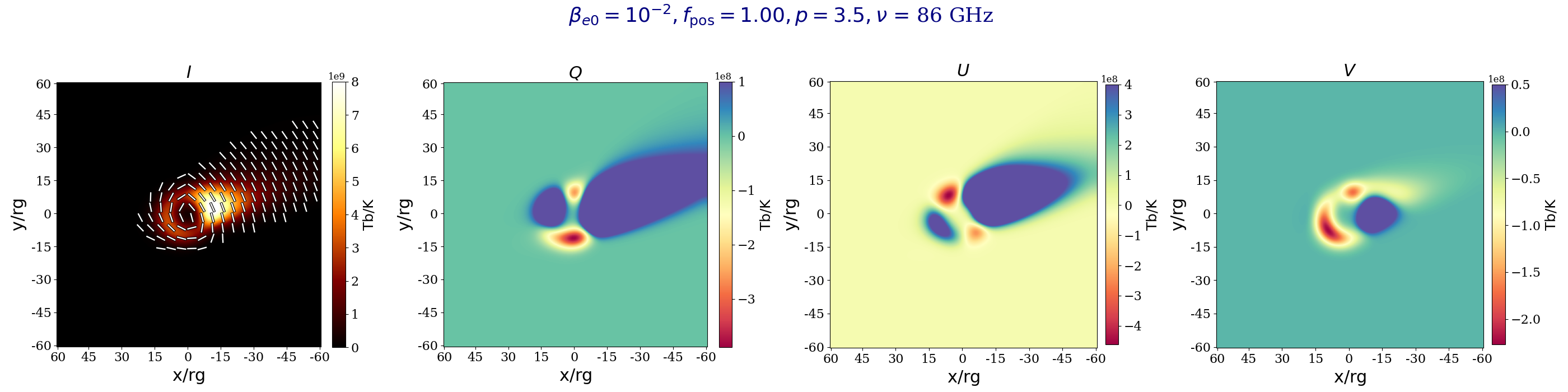}
\caption{86 GHz Stokes maps for the fiducial M87* jet model with  $\beta_{\mathrm{e0}}=10^{-2}$ and $p$ = 3.5. From Top to Bottom, we vary the fraction of positrons as $\beta_{\rm e0}=0.0,0.01,0.1,1.0$. All images are shown in a linear scale and are blurred with a circular Gaussian of FWHM 20 $\mu$as.
} \label{StokesMapsBetaE010Em2Pnth3Pt2Theta20fPos0AndPt5And1_86}
\end{figure*}
%%%%%%%%%%%%%%%%%%%%%%%%%%%%%%%%%%%%%%%%%%%%%%%%%%

%%%%%%%%%%%%%%%%%%%%%%%%%%%%%%%%%%%%%%%%%%%%
\begin{figure*}
\center
\includegraphics[width=0.99\textwidth]{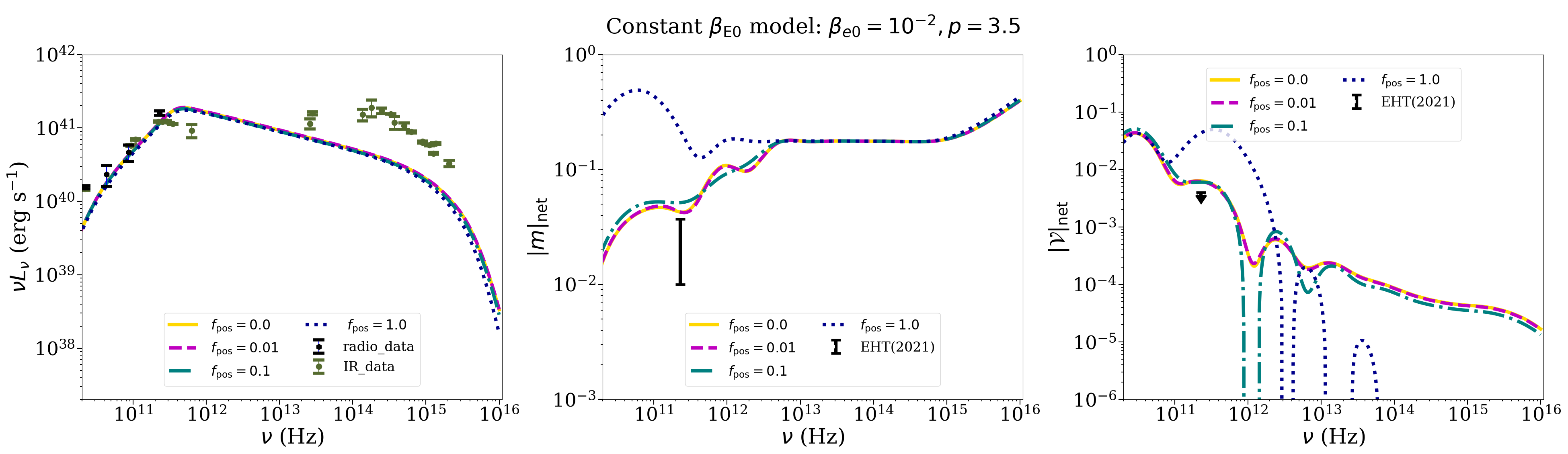}
\caption{ Total intensity spectral energy distribution (SED). (Left) for our best-bet M87* jet models with $f_{\mathrm{pos}} = (0.0, 0.01, 0.1)$ and for the best-bet model modified with a maximum pair fraction $f_{\mathrm{pos}} = 1.0$. Spectrum of the unresolved fractional linear polarization $|m|_{\rm net}$ (Center), compared with the EHT 2021 allowed range of 1.0-3.7\% \citep{Akiyama:2021fyp}. Spectrum of the unresolved fractional circular polarization $|v|_{\rm net}$ (Right). Note that none of our models satisfy the EHT constraints within $1\sigma$, but the fractional linear and circular polarization for our best-bet models are within $1.5\sigma$ of the EHT constraints. In contrast, both linear and circular polarization fractions are significantly above the current observational constraints for the full pair plasma case $f_{\rm pos}=1$. }
\label{best-bet-spectra}
\end{figure*} 
%%%%%%%%%%%%%%%%%%%%%%%%%%%%%%%%%%%%%%%%%%%% 
\subsection{Impact of $f_{\mathrm{pos}}$ on 86 GHz images}
\label{VaryingFrequencyFaraday}
So far, we have only presented Stokes maps at a single frequency of 230 GHz. 
In Figure \ref{StokesMapsBetaE010Em2Pnth3Pt2Theta20fPos0AndPt5And1_86}, we present Stokes maps for the M87* jet model at the frequency of $\nu = 86$ GHz for different values of the positron fraction. We notice several trends in the 86 GHz images w/r/t our 230 GHz images: 

$\bullet$ The intensity distribution at 86 GHz is more extended (as the plasma becomes optically thick) and the EVPA patterns become more ordered and radial than at 230 GHz.

$\bullet$ The linear polarization no longer peaks at the position of the black hole (as emission is optically thick there), but rather peaks farther along the jet.  We may see a similar effect in M87* \citet{Hada2016}.

$\bullet$ The simple quadrupolar pattern in linear polarization azimuthal variation observed at 230 GHz is no longer present, and we see a more complex pattern in the most polarized region downstream from the black hole 

$\bullet$ The 86 GHz circular polarization patterns remain ordered and bilaterally anti-symmetric in the non-pair plasma case. In the pair plasma case, the circular polarization peaks at the position of the black hole, whereas for the ionic-dominated cases it peaks further down the jet. 

The 86 GHz jet linear and circular polarization maps are distorted relative to the 230 GHz cases such as Fig. \ref{StokesMapsBetaE010Em2Pnth3Pt2Theta20fPos0AndPt5And1} further from the core, as the $\tau=1$ surface of unit optical depth expands outward (note the absorption coefficients for intensity and linear polarization scale as $\nu^{-\frac{p}{2}-2}$ and for circular polarization as $\nu^{-\frac{p}{2}-\frac{5}{2}}$). 

\subsubsection{Impact of $f_{\mathrm{pos}}$ on polarized spectra}
\label{best-bet-spectra}
Next, we focus on the total intensity SED and the spectra of fractional linear polarizations $|m|_{\rm net}$ and $|v|_{\rm net}$ for our best-bet models exploring the impact of $f_{\mathrm{pos}}$ in the polarized spectra. In producing spectra, we have limited the field of view (FOV) to 120 $r_{g}$. Consequently, these spectra do not contain extended emission on large scale and cannot account for the low-frequency slope in total intensity below $\nu=10^{10}$ Hz. 

Fig. \ref{best-bet-spectra} presents the total intensity spectral energy density (SED), the spectrum of the unresolved fractional linear polarization and the spectrum of the unresolved fractional circular polarization for our best-bet model. We also present spectra for the case with $f_{\mathrm{pos}} = 1$ for comparison. Included in each plot are the different observational constraints for the total intensity SED and for the linear and circular polarization discussed in Section~\ref{sec:observations}. We see that while the best-bet models are relatively close to the current polarimetric observational constraints, models with $f_{\mathrm{pos}} = 1$ sit well above the current EHT constraints are thus fully ruled out.

\subsection{RIAF Model Images and Spectra \label{subsec:RIAFImagesAndSpectra}}
Here we turn our attention to the RIAF model images and spectra for Sgr A*. As we discussed above, both because current EHT polarimetric constrains for Sgr A* are less robust than for M87* and because these models have been well-fit to available data in \citet[e.g.][]{Broderick2016}, we fix most parameters and only vary plasma $\beta$ and $f_{\rm pos}$ in the present work.

\subsubsection{RIAF Model: Impact of $f_{\mathrm{pos}}$ on 230 GHz images}
First, we consider the Stokes maps for our fiducial RIAF model. The parameters we use are $\beta= 10^{-2}$, $f_{\mathrm{nth}} = 0.01$, $T_e = 1.5 \times 10^{11} K$, $\gamma_{\mathrm{min}} = 10$, $\gamma_{\mathrm{max}} = 10^6$, and  $p$ = 2.8. We fix the inclination at $i = 60^\circ$ and orient the image at a position angle of $\xi = 156 ^\circ$  East of North. The BH spin is kept fixed at $a$ = 0.1 \citep{Broderick2016}. We blur all image with a circular Gaussian beam to a resolution of 15$\mu$as.

Fig. \ref{Stokes-RIAF-fnthPt001-Te1Pt5E11-PNTH2Pt8-betaEq10-fpos0AndPt5And1} shows the effects of increasing positron content in the RIAF models with all other parameters kept fixed. All images show a polarized crescent with a radial EVPA distribution. The linear polarization images $Q$ and $U$ exhibit azimuthal variation in two peak-valley cycles 
-- these features can be naturally interpreted as arising from the toroidal field geometry we fix in the model. Note that special relativistic beaming causes the approaching half of the disk to have a pronounced brightness enhancement. 
The addition of positrons result in a more ordered, more radial EVPA pattern. As in our exploration of M87* jet models, this increases in linear polarization with increasing positron fraction occurs because Faraday rotation (and thus beam depolarization) is suppressed \citep[e.g.][]{Jimenez2018}

In reality, a pair plasma RIAF accretion flow model is un-physical, as some ions are required to maintain the disk virial temperature. Furthermore, both pair-production scenarios we recap in  Section~\ref{sec:estimates} predict a polarization fraction $f_{\rm pos}\approx0$ for Sgr A*. Our results indicate that the addition of a relatively small population of positrons will not significantly modulate the polarized emission at 230 GHz. 
%%%%%%%%%%%%%%%%%%%%%%%%%%%%%%%%%%%%%%%%%%%%%%%%%%%%
\begin{figure*}
\center
\includegraphics[width=1.0\textwidth]{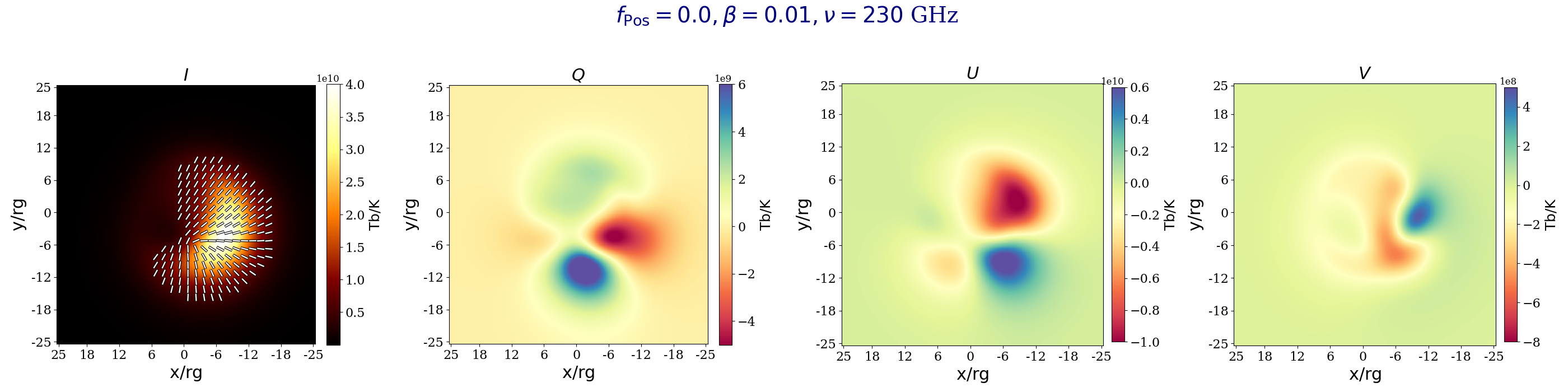}
\includegraphics[width=1.0\textwidth]{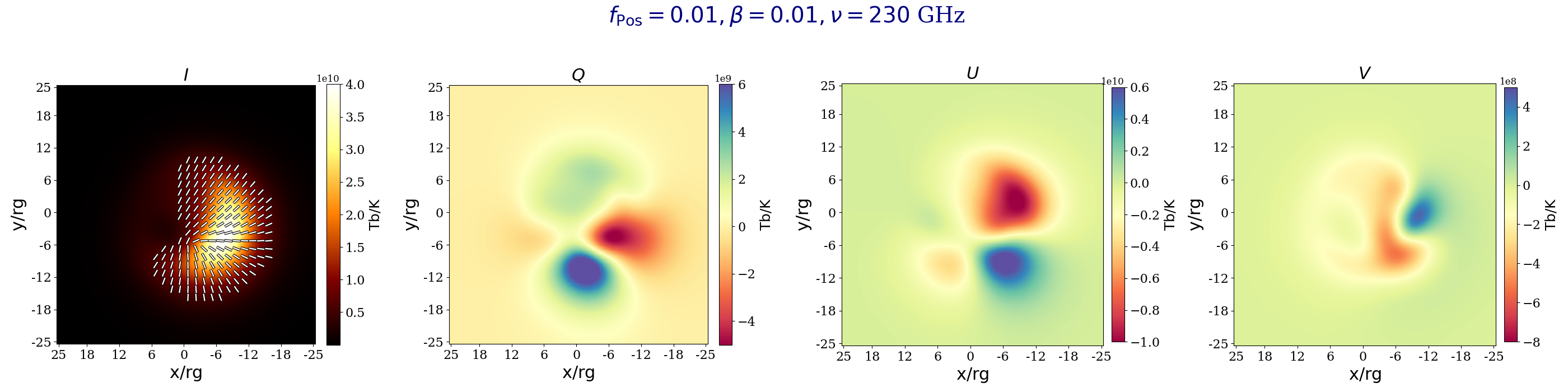}
\includegraphics[width=1.0\textwidth]{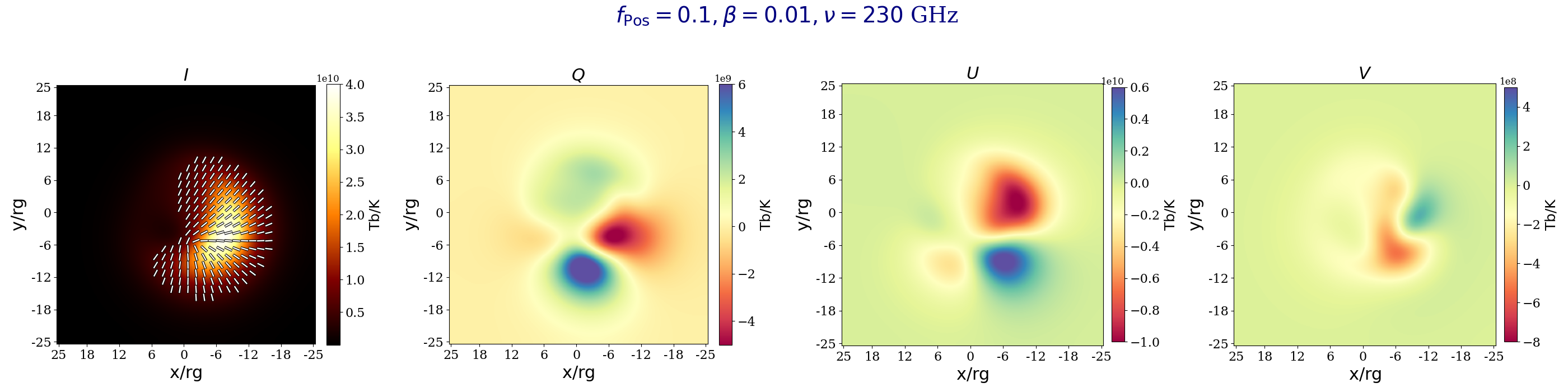}
\caption{Stokes maps for the Sgr A* RIAF models in Section~\ref{subsec:RIAFImagesAndSpectra}. Before display, the images were blurred with a circular Gaussian of 15$\mu$as FWHM. EVPA ticks in the left panel are shown above 5\% of the maximum linear polarization fraction. The models shown have 
with $\beta=10$, $f_{\mathrm{nth}} = 0.01$, $T_e = 1.5 \times 10^{11} K$ and  $p$ = 2.8. From Top to Bottom, we vary the positron fraction as $f_{\rm pos}=0,0.01,0.1,1.0$.}
\label{Stokes-RIAF-fnthPt001-Te1Pt5E11-PNTH2Pt8-betaEq10-fpos0AndPt5And1}
\end{figure*}
%%%%%%%%%%%%%%%%%%%%%%%%%%%%%%%%%%%%%%%%%%%%%%%%%%

 %%%%%%%%%%%%%%%%%%%%%%%%%%%%%%%%%%%%%%%%%%%%%%%%%%%%
\begin{figure*}
\center
\includegraphics[width=1.0\textwidth]{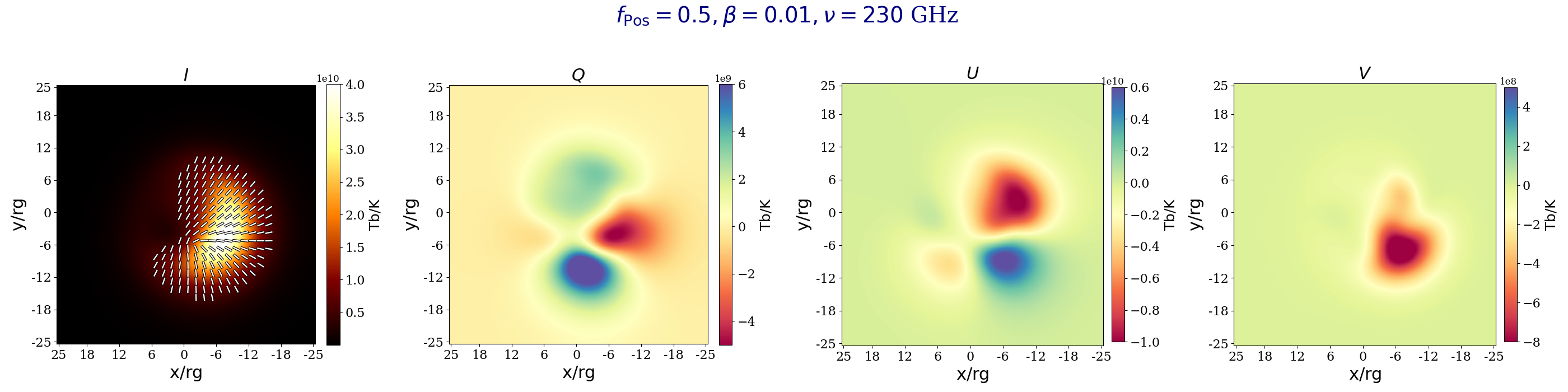}
\includegraphics[width=1.0\textwidth]{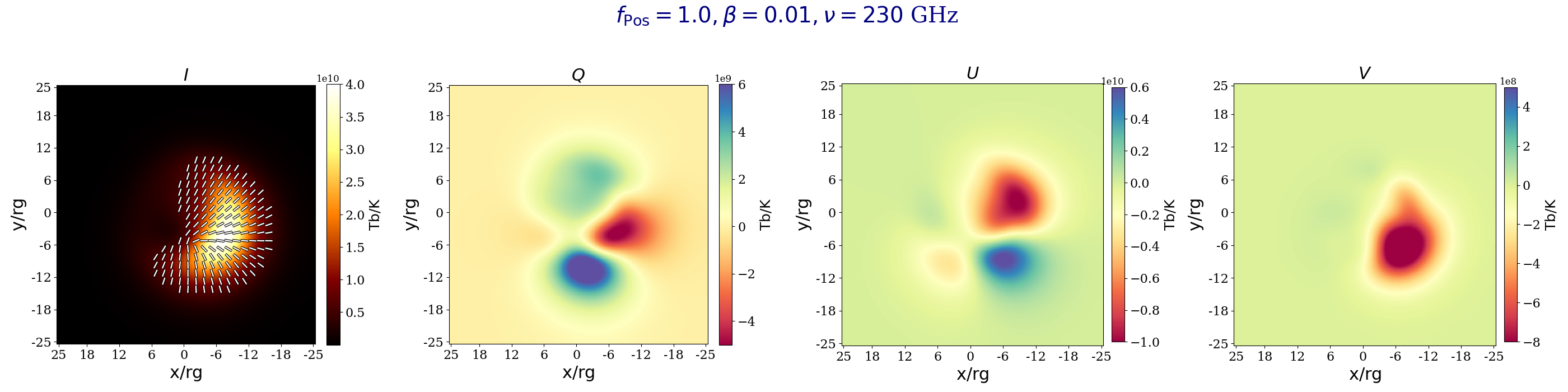}
\caption{Stokes maps for the Sgr A* RIAF models in Section~\ref{subsec:RIAFImagesAndSpectra}. Before display, the images were blurred with a circular Gaussian of 15$\mu$as FWHM. EVPA ticks in the Left panel are shown above 5\% of the maximum linear polarization fraction. The models shown have 
with $\beta=10$, $f_{\mathrm{nth}} = 0.01$, $T_e = 1.5 \times 10^{11} K$ and  $p$ = 2.8. From the Top to Bottom, we vary the positron fraction as $f_{\rm pos}=0,0.01,0.1,1.0$.
}
\label{Stokes-RIAF-fnthPt001-Te1Pt5E11-PNTH2Pt8-betaEq10-fpos0AndPt5And1}
\end{figure*}
%%%%%%%%%%%%%%%%%%%%%%%%%%%%%%%%%%%%%%%%%%%%%%%%%%
 \subsubsection{RIAF Model: impact of $f_{\mathrm{pos}}$ on polarized spectra}
Having presented the Stokes maps for RIAF models, below we discuss their total intensity SEDs and fractional polarization spectra. Figure \ref{PolSpectralFluxRIAFfnthEq10Em2BetaEq10Em1And1And10Pnth2Pt8Theta45fPos0AndPt5And1} presents the total intensity SED (left), the unresolved fractional linear (center) and circular (right) polarizations for our fiducial models with $\beta = 10^{-2} $ and with $f_{\mathrm{pos}} = (0.0, 0.01, 0.1)$. Added to each panel, we also present the case with $f_{\mathrm{pos}} = 1.0$. In each panel, we have also added the current observational constraints as they were mentioned in Sec. \ref{sgrA-observation}.

$\bullet$ First considering the total intensity SED, it is evident that different models are degenerate in the radio band (dominated by thermal self-absorption) and in the IR (dominated by power-law emission). Changing the plasma-$\beta$ (and thus the magnetic field strength), shifts the position of the synchrotron peak in the intermediate range $5 \times 10^{11} \leq \nu/\mathrm{Hz} \leq 5 \times 10^{13}$. As we saw in the M87* jet model, smaller values of $\beta$ (larger field strengths) peak at higher frequencies. Models at the same $\beta$ show only small differences in the transition region from radio to IR frequencies with different values of $f_{\rm pos}.$

$\bullet$ For the unresolved fractional linear polarization, at low frequencies, $\nu \leq 10^{11} $ Hz, models with higher positron fractions are more polarized. The enhancement in $|m|_{\rm net}$ with large $f_{\rm pos}$ is again due to the suppression of Faraday rotation in models with more positron content. As in M87* models, internal Faraday rotation is the primary source of depolarization in such models, as the intrinsic polarization pattern is rotated by different amounts on different lines-of-sight. While Faraday rotation peaked in the submm in our M87* jet models, however, here it is most significant at frequencies $<10^{11}$Hz.

$\bullet$ Finally, we see the same pattern in the unresolved fractional circular polarization as we observed in the M87* jet model. The circular polarization spectrum drops significantly in the pair plasma toward higher frequencies, owing to the inefficiency of the Faraday conversion in this limit (note $\rho_V\propto\nu^{-2}$) and the suppression of all intrinsic Stokes $V$ emission in a pair plasma. 

%%%%%%%%%%%%%%%%%%%%%%%%%%%%%%%%%%%%%%%%%%%%
\begin{figure*}
\center
\includegraphics[width=1.0\textwidth]{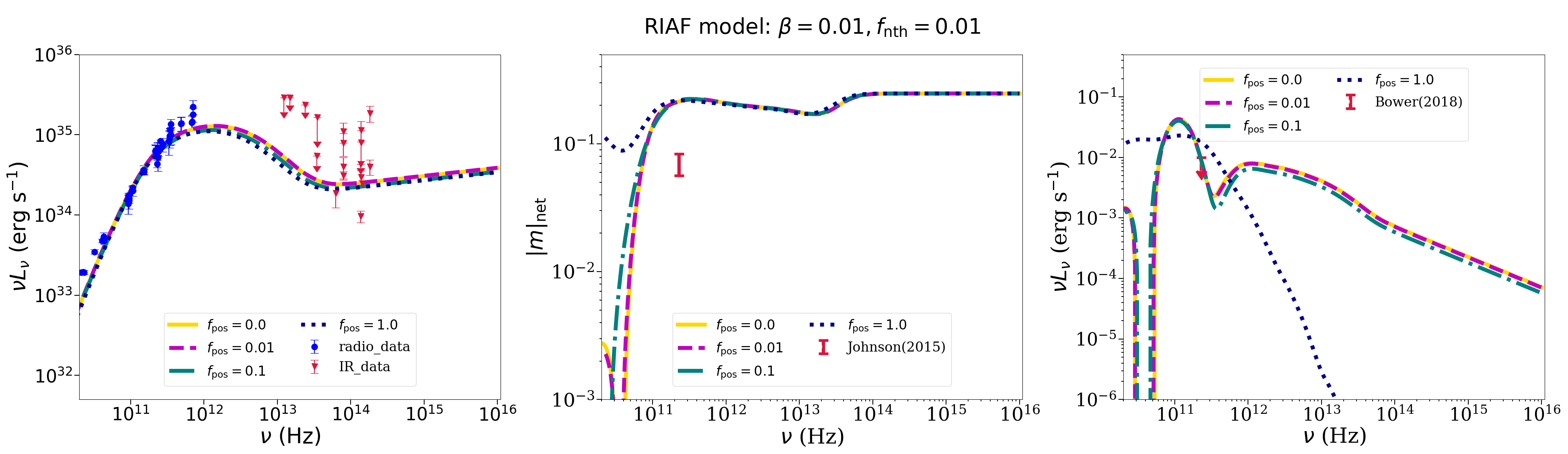}
\caption{
Total intensity spectral energy distribution (SED) (Left) for our best-bet Sgr A* RIAF models with $f_{\mathrm{pos}} = (0.0, 0.01, 0.1)$ and for the best-bet model modified with a maximum pair fraction $f_{\mathrm{pos}} = 1.0$.   Spectrum of the unresolved fractional linear polarization $|m|_{\rm net}$ (Center), compared with the \cite{2015Sci...350.1242J}. Spectrum of the unresolved fractional circular polarization $|v|_{\rm net}$ (Right).
We fix the fraction of nonthermal-to-thermal particles as $f_{\mathrm{nth}}=10^{-2}$ and the nonthermal power-law slope $p=2.8$. }
\label{PolSpectralFluxRIAFfnthEq10Em2BetaEq10Em1And1And10Pnth2Pt8Theta45fPos0AndPt5And1%fig:PolSpectralFluxBetaE010Em2BetaCrit10Em1And10Pnth3Pt2Theta20fPos0AndPt5And1
}
\end{figure*}
%%%%%%%%%%%%%%%%%%%%%%%%%%%%%%%%%%%%%%%%%%%%

%%%%%%%%%%%%%%%%%%%%%%%%%%%%%%%%%%%%%%%%%%%%%%%%%%%%%
\section{Conclusions and Future Directions}
\label{sec:summary}
\subsection{Conclusions}
In this paper, we have considered the effects of adding a non-zero positron fraction to models of M87* and Sgr A*. Positrons may be generated by the Breit-Wheeler process in LLAGN, either in a high-efficiency spark-gap electron acceleration and inverse Compton cascade scenario \citep{Broderick2015}, or in a 
low-efficiency "pair drizzle"  scenario \citep{Moscibrodzka2011,Wong2021}. Both scenarios can give non-negligible positron-to-electron fractions in the range $f_{\rm pos}\sim10^{-3}-1$ in M87*; however, both predict essentially zero positron fraction in Sgr A*. While we take these results from the literature as motivating values for our choice of positron fractions explored in this work, we extend the explored range of values in both the Sgr A* and M87* models we consider to examine the observational effects of a nonzero pair fraction in different source models in a way which is agnostic to the production mechanism. 

The first model we considered was for the M87* jet. We abstracted a GRMHD simulation into a semi-analytic model of a Poynting flux jet populated by non-thermal electrons and positrons. Fixing the flux at 230 GHz to its observed values from the recent observations of M87*, we found the magnetic field scale $B_0$ for each model, varying the distribution function slope $p$, the electron-to-magnetic pressure ratio $\beta_{\rm e0}$, and the positron fraction $f_{\rm pos}$. Analyzing the reduced $\chi^2$ of the resulting total intensity SEDs and the 230 GHz constraints on the polarized structure, we selected a reasonable fiducial model. We investigated the impact of changing $\beta_{\rm e0}$ and $f_{\rm pos}$ on the Stokes maps and the spectra. We found that increasing the positron content in the constant $\beta_{\rm e0}$ model increases both the 230 GHz linear polarization fraction (from decreased Faraday rotation and beam depolarization) and the circular polarization fraction (from increased Faraday conversion). As a consequence, the pair content in the M87* 230 GHz emitting region is unlikely to exceed $f_{\rm pos}\approx10^{-2}$.  At 86 GHz, our jet models are most polarized downstream from the core, with a distinct polarization pattern from the 230 GHz images. 
    
We next considered a RIAF model of Sgr A*'s accretion flow \citep{Broderick2009}. This geometric model describes a torus with power-law distributions of temperature and number density height and a coherent, toroidal $B$-field (poloidal and toroidal). We put $1\%$ of electrons and positrons in a power law component, 
but thermal electrons and positrons supply most of the emission. 
Throughout out analysis, we fixed the best-fit parameters from \citep{Broderick2016} and investigated the impact of varying plasma-$\beta$ and the positron fraction $f_{\mathrm{pos}}$ on the Stokes maps at 230 GHz and on the polarized spectra. We show that, we can account for the total intensity and total flux spectra with $1\%$ of the number density in power law electrons and/or positrons. Changing $\beta$ does not alter the total intensity SED significantly. The key discriminant among degenerate compositions is in the polarization, where decreased Faraday rotation from higher pair fractions result in higher linear polarization fractions at low frequencies. On the other hand, the increased Faraday rotation from lower pair fractions at low frequencies leads to somewhat smaller linear polarization at low frequencies. 

In both models, the overall changes in both the linear and circular polarization with low values of $f_{\rm pos}$ are small. The analysis in this work suggests that large positron fractions $f_{\rm pos}\gtrsim0.5$ have a significant effect on the polarized SEDs and 230 GHz images from both model classes, and the application of EHT constraints from \citep{Akiyama:2021fyp} suggests that the 230 GHz M87* emission region is predominantly ionic. However, these results also suggest that current constraints are unlikely to rule out or distinguish between pair fractions $\ll 50$\%. 

\subsection{Future Directions}
The analysis in this paper is an initial foray into examining the effects on non-zero positron content on realistic physical models of the 230 GHz images and broadband spectra from jet/accretion flow models of Sgr A* and M87*. We plan to extend this analysis in several directions in a future work:

\begin{enumerate}
\item A major limitation of the present analysis is the assumption of a constant positron-to-electron fraction throughout the source. Pair-drizzle models, in contrast, predict azimuthal structure in $n_{e^+}$, with the pair fraction peaking at the poles in the jet region \citep{Moscibrodzka2011, Wong2021}. In a forthcoming work, we will extend our models to include spatially varying parameterizations for $f_{\rm pos}.$

\item The current work has focused on introducing the models we consider and presenting a few best-bet models and qualitative trends in the models when increasing $f_{\rm pos}$. In a forthcoming work, we will narrow this on the preferred parameter space determined here and provide a more complete fit to the observational (total intensity and polarimeteric) multi-frequency data, determining fitted values for model parameters like $p$, $\beta_{\rm e0}$ and critically $f_{\rm pos}$ with associated uncertainties from an MCMC posterior sampler.  

\item We have chosen a particular  semi-analytic jet model for M87* derived from a GRMHD simulation \citep{Anantua2020a} for our M87* jet investigation. The field structure in this model is fixed and not varied in our analysis. Similarly, in our main RIAF model from \cite{Broderick2006b}, we fix the field orientation. The fixed field orientations determine our underlying EVPA and Stokes $V$ patterns in both models. 

\item In this work, we consider only a pure jet model for M87* and a pure disk model for Sgr A*. We chose these models to bracket the parameter space and explore the effects of positrons on images of two well-known source morphologies. However, a face-on jet model could explain both the observed Sgr A* SED and 230 GHz image size \citep{Issaoun2019}, and M87*'s 230 GHz emission may originate in the equatorial plane from a MAD disk \citep{Akiyama:2019fyp,Chael2021}.
In realistic near-horizon accretion flow simulations, both disk and jet emissions are present and can contribute to the 230 GHz image. A hybrid jet+accretion flow model may be a better choice to model both sources than the limiting cases which we explore here. 

\item Both our Sgr A* and M87* models are static with smooth plasma and field structure, but we know both sources are turbulent and dynamical and feature fine-scale structure. In M87* in particular, turbulent magnetic fields can help depolarize the net emission, which may bring our 230 GHz results more comfortably in line with observations
(though Faraday depolarization is the dominant effect, see \citealt{Jimenez2018}). 
We plan to extend our models by adding a post-processing parameterization for turbulent structure.

\item Finally, we will incorporate models for non-zero positron fractions (with spatial dependence) into radiative post-processing of full GRMHD simulations. Adding positrons to GRMHD images will naturally allow us to explore the effects of positrons on different emitting regions (jet vs disk) as well as the effects of fine-scale turbulence and source variability in our models. 
By adding a prescription for nonzero $f_{\rm pos}$ to the production of ray-traced images from GRMHD simulations such as presented in \citep{Akiyama:2019fyp,Akiyama:2021fyp}, we can also constrain the positron fraction directly from the simulation images following the procedures in \citep{Akiyama:2021fyp} for model evaluation. 

\end{enumerate}

\section*{Acknowledgements}
It is a great pleasure to thank  Charles Alcock, Charles Gammie, Michael Johnson, Geoff Bower,  Lars Hernquist, Angelo Ricarte, David Spergel and Craig Walker for the very insightful conversations. R.E. and R.A. acknowledges the support by the Institute for Theory and Computation at the Center for Astrophysics. This work was supported in part by the Black Hole Initiative at Harvard University, which is funded by grants from the John Templeton Foundation and the Gordon and Betty Moore Foundation. The computations in this paper were run on the FASRC Cannon cluster supported by the FAS Division of Science Research Computing Group at Harvard University. RA acknowledges the Center  for Computational Astrophysics funded by the Simons Foundation. Support for this work was provided by NASA through the NASA Hubble Fellowship grant HST-HF2-51431 awarded by the Space Telescope Science Institute, which is operated by the Association of Universities for Research in Astronomy, Inc., for NASA, under contract NAS5-26555.

\textit{Software:} {\tt GRTRANS} \citep{Dexter2016}, matplotlib \citep{2007CSE.....9...90H}, numpy \citep{2011CSE....13b..22V}, scipy \citep{2007CSE.....9c..10O}.

%\clearpage
%\newpage
\bibliography{main}{}
\bibliographystyle{aasjournal}

\appendix
\section{Estimate of $\beta_{\rm e}$ for M87* in a BZ Jet Interpretation}\label{EstimatingJetBeta}
In this section, we provide estimates of $\beta_{\rm e0}$ in the submm emission region of M87* that we use to motivate our choices for the jet model explored in Section~\ref{sec:modelspace}. We make use of the the partial electron pressure, \citep{Blandford:2017pet,Anantua2020a}:
\begin{equation}
    \tilde{P}_{\rm e}=
     \frac{1}{3}\gamma^{'2}N_{e^-}(\gamma')m_{\rm e}c^2,
\end{equation}
which is the contribution to the total pressure from relativistic electrons emitting within an octave of the observed frequency $\nu$ (where primes denote the fluid co-moving frame, so $\gamma'$ is the local electron Lorentz factor that produces the observed emission at $\nu$,  and $N_{\rm e^-}$ is the electron distribution function per unit Lorentz factor). In the context of mm-emitting electrons, we can relate the partial pressure to the observed luminosity $L_\nu$ and local cooling time:
\begin{equation}
     \tilde{P}_{\rm e} = \frac{1}{3}\frac{L_\nu t_\mathrm{cool}}{V_\mathrm{emit}}
\end{equation}\label{PeEstimate}
where $V_\mathrm{emit}$ is the volume of the emitting region and $t_\mathrm{cool}$ is the synchrotron cooling time: 
\begin{equation}
    t_\mathrm{cool}=\frac{10}{\gamma}\left(\frac{10^{-4}}{B^2}\right)^2\mathrm{yr} 
\end{equation}
If we neglect the effects of red/blue-shift, the emitting Lorentz factor $\gamma'(\nu)$ is:
\begin{equation}
    \gamma'^2=\frac{m_{\rm e}}{3\pi e B}\nu.
\end{equation}
Note we have modeled the emission in terms of the total pressure in the main text. Here, we seek an independent estimate of characteristic $\beta_{\rm e0}$ using only the partial pressure contribution from emitters that contribute to the observed frequency.

To estimate $\beta_{\rm e0}$ in different emission regions along the M87* jet, we use the jet geometry given by \cite{Hada2013}: the cylindrical radius varies with height as $s\propto z^{0.56}$ out to a de-projected distance $z\approx 200r_{\rm g}$. The synchrotron luminosity $L_\nu$ can be deduced from the spectrum from \cite{Prieto2016}. Then, we can estimate $\tilde{P}_{\rm e}$ and $\beta_\mathrm{e}=\tilde{P}_{\rm e}/P_B$ if we have an estimation for the B-field along with the jet. For this, we use the same jet model in Eq.~\ref{eq:bfield}, originally set in \cite{Anantua2020a} based on the MB09D simulation from \cite{McKinney2012}. Although in the main text we set the scaling factor $B_0$ based on which value gives the desired 230 GHz flux, here we make estimates using a fixed field strength set by using the jet magnetic flux (Eq.~\ref{eq:flux}) $\Phi_{H,\mathrm{M87}}\approx B_0r^2_{\rm g} \approx 10^{34}\  \mathrm{Mx}$ \citep{Anantua2020a} (conversions between the cgs scaling factors used in this paper and the mks unit scaling factors in \citet{Anantua2020a} are given in Table \ref{Tab:CodeToPhysicalUnits}). Note this jet flux of $10^{34}$ Mx produces a jet much more powerful than the EHT lower limit of $L_\mathrm{jet}\gtrsim 10^{42}$ erg s$^{-1}$ used in \citet{Akiyama:2019fyp}.

%%%%%%%%%%%%%%%%%%%%%%%%%%%%%%%%%%%%%%%%
\begin{table*}
%\linewidth
\centering
\caption{Code-to-physical unit conversions used in \cite{Anantua2020a}. The mapping from these to the magnetic field conversion factors we use here to conserve energy between synthetic and observed M87* intensity maps with target flux 1.5 Jy is multiplication the magnetic field scale $B_0$, as given in Table \ref{chi2-constant-beta}.
In \cite{Anantua2020a}, there are prefactors $\Phi_{26}10^{26}$ and $r_{13}10^{13}$ where $\Phi_{26}\approx1\approx r_{13}$  that link code units to physical units (in cgs) that normalize the horizon flux to 10$^{34}$ Mx (by comparison, our flux normalization $B_0$ in this paper gives a total intensity of 1.5 Jy). Including both of these scaling methods allows the reader to go back and forth between the two papers. 
}
	\label{Tab:CodeToPhysicalUnits}
	\begin{tabular}{lccr} 
		\hline
		Code Unit & 
        mks & cgs  \\
     &
     & 
     \\
     		\hline
		Flux threading (one hemisphere %&  &  & \\
         % & $1.03897\cdot 10^{20}$ 
          & $\Phi_{26}10^{26}$
          Wb & 
          $\Phi_{34}10^{34}$
          Mx
          \\
          of) the horizon ($\Phi_B$) &&\\
		\hline
		 Magnetic Field Near Horizon ($B_\mathrm{scale}$)%($r_g=r_{13}10^{13}$m) 
		 &  
        %&  
        & \\
        $\frac{\Phi_B}{(\mathrm{Length\ Unit})^2}$& 
        %$9.66833\cdot 10^{20}$ & 
        $\frac{10^{26}\Phi_{26}\mathrm{Wb}}{(r_{13}10^{13}\mathrm{m})^2}=\frac{\Phi_{26}}{r_{13}^2}\mathrm{T}$%New (f,betaTotCrit);$-1.48619\cdot 10^{-7}$%%Old (f,betaCrit)
        & $\frac{10^{34}\Phi_{34}\mathrm{Mx}}{(r_{15}10^{15}\mathrm{cm})^2}=\frac{\Phi_{34}}{r_{15}^2}10^4\mathrm{Gs}$
      \\
        \hline
		Pressure Near Horizon &  &  & \\
         $\left(\frac{\Phi_B}{(\mathrm{Length\ Unit})^2}\right)^2$   & 
         %$1.90238\cdot 10^{19}$ & 
         $\frac{1}{2\mu_0}\left(\frac{\Phi_{26}}{r_{13}^2}\mathrm{T}\right)^2=
         \frac{1}{8\pi\cdot 10^{-7}}\frac{\Phi_{26}^2}{r_{13}^4}$ Pa& $\frac{1}{8\pi}\left(\frac{\Phi_{34}}{r_{15}^2 10^4}\mathrm{Gs}\right)^2=
         \frac{10^8}{8\pi}\frac{\Phi_{34}^2}{r_{15}^4}$ Ba%40437
         \\
        \hline
	\end{tabular}\label{CodeToPhysical}
\end{table*}
%%%%%%%%%%%%%%%%%%%%%%%%%%%%%%%%%%%%%%
Using the values of emitting region height and size from \citet{Hada2013} and the powerful $B$-field estimates from the $10^{34}$ Mx jet model \citep{Anantua2020a}, we estimate values of the synchrotron cooling time (at the boundary of the \citet{Hada2013} optically thick core) as $\sim 2.6\times10^4$s, $4.3\times10^5$s, $6.6\times10^6$s, and $3.9\times10^8$s for $\nu = 230$ GHz, 86 GHz, 43 GHz, and 22 GHz respectively. Then we estimate  $\beta_{\rm e0}$ in these emission regions as
$\beta_{\rm e0} \approx \{ (230\ \mathrm{GHz}, 1.8\times 10^{-5}), (86\ \mathrm{GHz}, 9.2\times10^{-6}), (43\ \mathrm{GHz}, 2.0\times 10^{-4}), (23\ \mathrm{GHz}, 2.2\times10^{-1})\}$.

The observational $\tilde{P}_\mathrm{e}$ estimation give us a ballpark  $10^{-6}\lesssim\beta_{e0}\lesssim10^{-2}$ which helps motivate our choice of parameter space for the constant $\beta_{\rm e0}$ jet model in in Section~\ref{sec:modelspace}.
 
\section{$\chi^2$ Tables}\label{ChiSqJetModelAppendix}
In this section, we present tables of reduced-$\chi^2$ values we compute by comparing model Stokes $I$ SED against the observed M87* or Sgr A* SEDs to select models. We present $\chi^2$ values considering only points in the radio-submm band ($\chi^2_{\rm radio}$, in the IR ($\chi^2_{\rm IR}$) and from the full SED data-sets ($\chi^2$). 

The $\chi^2$ values for Constant $\beta_{\rm e0}$ jet models for M87* models are given in Table \ref{chi2-constant-beta}.
The $\chi^2$ values for the Sgr A* RIAF model can be found in Table \ref{RIAF-Model}.
Note that in the RIAF model we consider a smaller parameter space than in the M87* jet model by only varying $\beta$ and $f_{\rm pos}$ and fixing other parameters to the fitted values from \citet{Broderick2016}.

%%%%%%%%%%%%%%%%%%%%%%%%%%%%%%%%%%%%%%%%%
\begin{table}
%\hspace*{-22em}
% \begin{adjustwidth}{-2cm}{}
\caption{Stokes I SED reduced $\chi^2$ values for different constant electron $\beta_{\rm e0}$ M87* jet models. In each section of five rows, we fix the value of the positron fraction $f_{\rm pos}$ and vary the electron distribution spectral slope $p\in\{2.5,2.8,3.0,3.2,3.5\}.$ The three sections of five columns each contain data for different values of $\beta_{\rm e0}\in\{10^{-6},10^{-4},10^{-2}\}$.
The best $\chi^2$ over $f_{\mathrm{pos}}, p $ and $\beta_{e0}$ for each segment of the spectrum are given in bold.
}
\makebox[ 0.83 \textwidth][c]{
\begin{tabular}{cc|ccccc|ccccc|ccccc}
\toprule
\hline
$f_{\mathrm{pos}}$ & $p$  & $\beta_{\mathrm{e0}}$  &  $B_0$($ 10^3$) & $\chi_{\mathrm{radio}}^2$  & $\chi_{\mathrm{IR}}^2$ & $ \chi^2_{\mathrm{tot}}$ & $\beta_{\mathrm{e0}}$  &  $B_0$($ 10^3$)   & $\chi_{\mathrm{radio}}^2$  & $\chi_{\mathrm{IR}}^2$ & $ \chi^2_{\mathrm{tot}}$ & $\beta_{\mathrm{e0}}$  &  $B_0$($ 10^3$) & $\chi_{\mathrm{radio}}^2$  & $\chi_{\mathrm{IR}}^2$ & $ \chi^2_{\mathrm{tot}}$ 
\\ \midrule
0.00 & 2.5 & $10^{-6}$ & 11.7  & 11.0 & $>10^4$  & $>10^4$ & $10^{-4}$ & 3.1 & 7.78 & $>10^3$ & $>10^3$ & $10^{-2}$ & 0.88 & 6.1 & $\gtrsim 10^3$ & $\sim 10^3$ \\ \hline
0.00 & 2.8 & $10^{-6}$ & 11.3  & 9.6 & $\gtrsim 10^3$  & $\sim 10^3$ & $10^{-4}$ & 3.2 & 7.4 & 610.2 & 319.2 & $10^{-2}$ & 0.93 & 5.5 & 93.0 & 50.7 \\ \hline
0.00 & 3.0 & $10^{-6}$ & 11.7  & 10.3 & 426.2  & 225.4 & $10^{-4}$ & 3.3 & 7.1 & 92.3 & 51.1 & $10^{-2}$ & 0.98 & 5.2 & 7.00 & 6.1 \\ \hline
0.00 & 3.2 & $10^{-6}$ & 12.1  & 10.4 & 82.4  & 47.6 & $10^{-4}$ & 3.5 & 7.3 & 13.1 & 12.3 & $10^{-2}$ & 1.07 &  5.6 & 1.2 & 3.3 \\ \hline
0.00 & 3.5 & $10^{-6}$ & 12.5  & 9.8 & 1.9  & 5.7 & $10^{-4}$ & 3.9 & 8.0 & 3.1 & 5.5 & $10^{-2}$ & 1.2 & 5.7 & 7.7 & 6.8
\\ \hline \hline
0.01 & 2.5 & $10^{-6}$ & 11.6 & 10.6 & $>10^4$  & $>10^4$ &  $10^{-4}$ & 3.1 & 7.88 & $>10^3$ & $>10^3$ &  $10^{-2}$ & 0.88 & 6.15 & $\gtrsim 10^3$ & $\sim 700$ \\ \hline
0.01 & 2.8 & $10^{-6}$ & 11.6  & 10.7 & $\gtrsim 10^3$  & $\sim 10^3$ & $10^{-4}$ & 3.2 & 7.3 & 591 & 309.5 & $10^{-2}$ & 0.92 & 5.4 & 80.7 & 44.3 \\ \hline
0.01 & 3.0 & $10^{-6}$ & 11.6 & 9.96 & 394.4  & 208.8 & $10^{-4}$ & 3.4 & 7.4 & 107.7  & 59.3 & $10^{-2}$ & 0.99 & 5.39 & 10.0 & 7.78 \\ \hline
0.01 & 3.2 & $10^{-6}$ & 12.2  & 10.92 & 92.96  & 53.35 & $10^{-4}$ & 3.5 & 7.29 & 13.18 & 10.34 & $10^{-2}$ & 1.07 & 5.51 & 1.19 & 3.27 \\ \hline
0.01 & 3.5 & $10^{-6}$ & 12.5  & 9.91 & 2.0  & 5.82 & $10^{-4}$ & 3.97 & 8.63 & 2.56 & 5.49 & $10^{-2}$ & 1.2 & 5.75 & 7.65 & 6.7 
\\ \hline \hline
0.1 & 2.5 & $10^{-6}$ & 11.3 & 10.45 & $>10^4$  & $>10^4$ & $10^{-4}$ & 3.1 & 7.82 & $>10^3$ & $>10^3$ &  $10^{-2}$ & 0.84 & 5.79 & $\gtrsim 10^3$ & 583 \\ \hline
0.1 & 2.8 & $10^{-6}$ & 10.99  & 9.35 & $\gtrsim 10^3$  & $\sim 870$ & $10^{-4}$ & 3.2 & 7.93 & 720 & 376.2 & $10^{-2}$ & 0.92 & 5.61 & 102.9 & 55.9 \\ \hline
0.1 & 3.0 & $10^{-6}$ & 11.3 & 9.72 & 373.2 & 197.7 & $10^{-4}$ & 3.3 & 7.34 & 104.4 & 57.53 & $10^{-2}$ & 0.95 & 5.2 & 6.57 & 5.9 \\ \hline
0.1 & 3.2 & $10^{-6}$ & 11.6  & 9.54 & 64.78  & 38.11 & $10^{-4}$ & 3.5 & 7.81 & 17.99 & 13.08 & $10^{-2}$ & 1.03 & 5.36 & 1.31 & 3.26 \\ \hline
0.1 & 3.5 & $10^{-6}$ & 12.2 & 9.7 & 1.93 & 5.68 & $10^{-4}$ & 3.8 & 7.88 & 3.17 & 5.44 & $10^{-2}$ & 1.18 & 5.54 & 8.18 & 6.91
\\ \hline \hline
0.5 & 2.5 & $10^{-6}$ & 10.2 & 9.6 & $>10^4$  & $\sim 10^4 $ & $10^{-4}$ & 2.8 & 8.0 & $> 10^3$ & $> 10^3 $ & $10^{-2}$ & 0.78 & 5.9 & $\gtrsim 10^3$ & 573 \\ \hline
0.5 & 2.8 & $10^{-6}$ & 10.2  & 9.39 & $ \gtrsim 10^3$  & $\sim 10^3$ & $10^{-4}$ & 2.9 & 7.6 & 624.6 & 326.7 & $10^{-2}$ & 0.83 & 5.4 & 70.5 & 39.0 \\ \hline
0.5 & 3.0 & $10^{-6}$ & 10.5  & 10.1 &  403.5 & 213.6 & $10^{-4}$ & 3.0 & 7.2 & 96.2 & 53.3 & $10^{-2}$ & 0.88 & 5.1 & 4.8 & 5.0 \\ \hline
0.5 & 3.2 & $10^{-6}$ & 10.9 & 10.2 & 78.8  & 45.7 & $10^{-4}$ & 3.2 & 7.5 & 14.8 & 11.3 & $10^{-2}$ & 0.98 & 5.5 & 1.1 & 3.2 \\ \hline
0.5 & 3.5 & $10^{-6}$ & 11.3  & 9.5 & 1.9  & 5.6 & $10^{-4}$ & 3.5 & 7.5 & 3.5 & 5.4 & $10^{-2}$ & 1.07 & 5.2 & 9.3 & 7.3 \\ \hline \hline 
1.0 & 2.5 & $10^{-6}$ & 9.4  & 6.5 & $> 10^4$  & $>10^3$ & $10^{-4}$ & 2.5 & 7.4 & $>10^3$ & $>10^3$ & $10^{-2}$ & 0.71 & 5.9 & $\sim 10^3$ & 405.7 \\ \hline
1.0 & 2.8 & $10^{-6}$ & 9.4  & 9.3 & $\gtrsim 10^3 $  & $\sim 10^3$ & $10^{-4}$ & 2.6 & 7.0 & 437.9 & 229.9 & $10^{-2}$ & 0.78 & 5.7 & 73.3 & 40.7 \\ \hline
1.0 & 3.0 & $10^{-6}$ & 9.8  & 9.7 & 379.3  & 201.0 & $10^{-4}$ & 2.7 & 6.8 & 63.7 & 36.2 & $10^{-2}$ & 0.83 & 5.5 & 5.5 & 5.5 \\ \hline
1.0 & 3.2 & $10^{-6}$ & 10.2  & 10.0 & 74.5  & 43.4 & $10^{-4}$ & 2.9 & 7.0 & 9.4 & 8.2 & $10^{-2}$ & 0.9 & 5.6 & 1.15 & 3.3 \\ \hline
1.0 & 3.5 & $10^{-6}$ & 10.4  & 8.8 & 1.9  & 5.2 & $10^{-4}$ & 3.3 & 7.7 & 3.2 & 5.3 & $10^{-2}$ & 1.0 & 6.0 & 8.6 & 7.2 \\ \hline
\bottomrule 
\end{tabular}
}
\label{chi2-constant-beta}
\end{table}
%%%%%%%%%%%%%%%%%%%%%%%%%%%%%%%%%%%%%%%%%

%%%%%%%%%%%%%%%%%%%%%%%%%%%%%%%%%%%%%%%%%
\begin{table}
\caption{Stokes $I$ SED reduced $\chi^2$ values for different Sgr A* RIAF models. In each block of 4 rows/5 columns, we fix the the positron fraction $f_{\mathrm{pos}} \in\{0.0, 0.01, 0.1\}$ and vary the plasma-$\beta\in\{10^{-2},10^{-1},1,10\}$. All other parameters are fixed to their values from \citet{Broderick2016}.
}
\makebox[ 0.83 \textwidth][c]{
\begin{tabular}{c|ccccc|ccccc|ccccc}
\toprule
\hline
$\beta$ & $f_{\mathrm{pos}}$  &  $n_{\rm th0}$($ 10^3$) & $\chi_{\mathrm{radio}}^2$  & $\chi_{\mathrm{IR}}^2$ & $ \chi^2_{\mathrm{tot}}$ & $f_{\mathrm{pos}}$  &  $n_{\rm th0}$($ 10^3$)   & $\chi_{\mathrm{radio}}^2$  & $\chi_{\mathrm{IR}}^2$ & $ \chi^2_{\mathrm{tot}}$ & $f_{\mathrm{pos}}$  &  $n_{\rm th0}$($ 10^3$) & $\chi_{\mathrm{radio}}^2$  & $\chi_{\mathrm{IR}}^2$ & $ \chi^2_{\mathrm{tot}}$
\\ \midrule
$10^{-2}$ & 0.0 & 314 & 1.61 & 4.51 & 1.89 & 0.01 & 313.6 & 1.6 & 4.48 & 1.88 & 0.1 & 283.1  & 1.78 & 4.8 & 2.53 \\ \hline
$10^{-1}$ & 0.0 & 893  & 2.02 & 4.95 & 2.30 & 0.01 & 893.4 & 2.0 & 4.93 & 2.28 &  0.1 & 832.3 & 2.14 & 5.06 & 2.42 \\ \hline
$10^{0}$ & 0.0 & 278 & 2.28 & 4.88 & 2.53 & 0.01 & 2785.3 & 2.25 & 4.86 & 2.50 & 0.1 & 2663.2 & 2.28 & 4.86 & 2.53 \\ \hline
$10^{1}$ & 0.0 & 8888  & 2.46 & 4.68 & 2.67 & 0.01 & 8888.2 & 2.44 & 4.66 & 2.65 & 0.1 & 8888.2  & 2.24 & 4.42 &  2.45 \\ \hline \midrule \hline 
$10^{-2}$ & 0.5 & 253 & 1.71 & 4.57 & 1.98 & 1.0 & 207 & 1.95 & 4.86 & 2.23 &-&-&-&-&- \\ \hline
$10^{-1}$ & 0.5 & 710 & 2.23 & 5.05 & 2.50 & 1.0 & 649 & 2.12 & 4.77 & 2.37 &-&-&-&-&- \\ \hline
$10^{0}$ & 0.5 & 3030 & 2.17 & 3.15 & 2.26 & 1.0 & 2053 & 2.33 & 4.62 & 2.55 &-&-&-&-&- \\ \hline
$10^{1}$ & 0.5 & 7423 & 2.46 & 4.54 & 2.66 & 1.0 & 6447 & 2.60 & 4.50 & 2.79 &-&-&-&-&- \\ \hline
\bottomrule 
\end{tabular}
}
\label{RIAF-Model}
\end{table}
%%%%%%%%%%%%%%%%%%%%%%%%%%%%%%%%%%%%%%%%%

\section{EHT 230 GHz Polarization Constraints}
\label{P-V-Observation}
In this section, we present tables of the 230 GHz polarimetric quantities - $|m|_{\rm net}$, $|v|_{\rm net}$, and $\langle |m|\rangle$ - that are used as constraints in the EHT model scoring analysis in \citet{Akiyama:2021fyp}. We present these quantities for both he M87* constant $\beta_{\rm e0}$ jet models and the Sgr A* RIAF models. For M87*, we present pass/fail values based on whether all three quantities satisfy the EHT constraints from M87* within 1.5$\sigma$. For Sgr A*, we do not present a Pass/Fail value, since we explore a more limited parameter space and because our preliminary data constraints on the resolved polarization $\langle |m|\rangle$ from \citet{Johnson2015} will soon be supplemented with a detailed analysis from resolved 2017 EHT Sgr A* images. Table~\ref{P-V-constant-beta} presents the 230 GHz polarimetric quantities for the M87* jet model, and Table~\ref{P-V-RIAF} present the 230 GHz polarimetric quantities for the Sgr A* RIAF. 

%%%%%%%%%%%%%%%%%%%%%%%%%%%%%%%%%%%%%%%%%
\begin{table}
\caption{
Table of 230 GHz polarimetric quantities $\left( |m|_{\mathrm{net}}, \langle |m| \rangle,
|v_{\mathrm{net}}| \right) $ for different M87* constant $\beta_{\rm e0}$ jet models. We also present an overall pass/rail score (P/F) based on whether all three quantities satisfy EHT 2017 constraints on the quantities within $1.5\sigma$ (note that none of the quantities fall within the $1\sigma$ range given in the constraint table from \citet{Akiyama:2021fyp}). 
In each section of five rows, we fix the value of the positron fraction $f_{\rm pos}$ and vary the electron distribution spectral slope $p\in\{2.5,2.8,3.0,3.2,3.5\}.$ The three sections of five columns each contains data for different values of $\beta_{\rm e0}\in\{10^{-6},10^{-4},10^{-2}\}$.
}
\makebox[ 0.83 \textwidth][c]{
\begin{tabular}{cc|ccccc|ccccc|ccccc}
\toprule
$f_{\mathrm{pos}}$ & $p$  & $\beta_{\mathrm{e0}}$  & 
$|m|_{\mathrm{net}}$ & $\langle |m| \rangle $ & 
$ |v_{\mathrm{net}}| $ &  \rm{Score}
& $\beta_{\mathrm{e0}}$  &
$|m|_{\mathrm{net}}$ & $\langle |m| \rangle $ & 
$ |v_{\mathrm{net}}| $ &  \rm{Score}
& $\beta_{\mathrm{e0}}$  & $|m|_{\mathrm{net}}$ & $\langle |m| \rangle $ & 
$ |v_{\mathrm{net}}| $ &  \rm{Score}
\\ \midrule
0.0 & 2.5 & $10^{-6}$ & 0.29 & 0.41 & 0.016  & F & $10^{-4}$ & 0.19 & 0.30 & 0.016 & F & $10^{-2}$ & 0.12 & 0.19 & 0.015 & F \\ \hline
0.0 & 2.8 & $10^{-6}$ & 0.26 & 0.38 & 0.016 & F & $10^{-4}$ & 0.16 & 0.25 & 0.017 & F & $10^{-2}$ & 0.08 & 0.14 & 0.012 & F \\ \hline
0.0 & 3.0 & $10^{-6}$ & 0.25 & 0.36 & 0.017 & F & $10^{-4}$ & 0.15 & 0.22 & 0.016 & F & $10^{-2}$ & 0.06 & 0.1 & 0.011 & F \\ \hline
0.0 & 3.2 & $10^{-6}$ & 0.25  & 0.36 & 0.017  & F & $10^{-4}$ & 0.13 & 0.20 & 0.014 & F & $10^{-2}$ & 0.05 & 0.079 & 0.01 & F \\ \hline
0.0 & 3.5 & $10^{-6}$ & 0.23 & 0.33 & 0.017 & F & $10^{-4}$ & 0.1 & 0.15 & 0.014 & F & $10^{-2}$ & 0.032 & 0.054 & 0.008 & P \\ \hline \hline
0.01 & 2.5 &  $10^{-6}$ & 0.29 & 0.41 &  0.016 & F &  $10^{-4}$ & 0.191 & 0.302 & 0.016 & F &  $10^{-2}$ & 0.117 & 0.195 & 0.015 & F \\ \hline
0.01 & 2.8 &  $10^{-6}$ & 0.267 & 0.379 & 0.016 & F &  $10^{-4}$ & 0.163 &  0.253 & 0.017 & F &  $10^{-2}$ & 0.087 & 0.141 & 0.012 & F \\ \hline
0.01 & 3.0 &  $10^{-6}$ & 0.25 & 0.359 & 0.016 & F &  $10^{-4}$ &  0.148 & 0.22 & 0.016 & F &  $10^{-2}$ & 0.067 & 0.099 & 0.011 & F \\ \hline
0.01 & 3.2 &  $10^{-6}$ & 0.258 & 0.360  & 0.017 & F &  $10^{-4}$ & 0.134 & 0.202 & 0.014 & F  &  $10^{-2}$ & 0.053 & 0.079 & 0.01 & F  \\ \hline
0.01 & 3.5 &  $10^{-6}$ & 0.234 & 0.327 &  0.017 & F &  $10^{-4}$ & 0.101 & 0.156 & 0.014  & F  &  $10^{-2}$ & 0.034 & 0.055 & 0.009 & P \\ \hline \hline
0.1 & 2.5 &  $10^{-6}$ & 0.296 & 0.413 &  0.015 & F &  $10^{-4}$ & 0.193 & 0.319 & 0.015 & F &  $10^{-2}$ & 0.131 & 0.221 & 0.016 & F \\ \hline
0.1 & 2.8 &  $10^{-6}$ & 0.271 & 0.387 & 0.0145 & F &  $10^{-4}$ & 0.175 & 0.266 & 0.016 & F &  $10^{-2}$ & 0.094 & 0.147 & 0.014 & F \\ \hline
0.1 & 3.0 &  $10^{-6}$ & 0.259 & 0.371 & 0.016 & F &  $10^{-4}$ & 0.153 &  0.233 & 0.017 & F &  $10^{-2}$ & 0.074 & 0.112 & 0.012 & F \\ \hline
0.1 & 3.2 &  $10^{-6}$ & 0.262 & 0.372 & 0.017 & F &  $10^{-4}$ & 0.146 & 0.217 & 0.015 & F &  $10^{-2}$ & 0.054 & 0.087 & 0.012 & F \\ \hline
0.1 & 3.5 &  $10^{-6}$ & 0.244 & 0.342 & 0.019 & F &  $10^{-4}$ & 0.106 & 0.171 & 0.015 & F &  $10^{-2}$ & 0.036 & 0.061 &  0.011 & P \\ \hline \hline
0.5 & 2.5 & $10^{-6}$ & 0.31  & 0.44 & 0.015  & F & $10^{-4}$ & 0.23 & 0.389 & 0.018 & F & $10^{-2}$ & 0.158 & 0.291 & 0.017 & F \\ \hline
0.5 & 2.8 & $10^{-6}$ & 0.30 & 0.42 & 0.015 & F & $10^{-4}$ & 0.201 & 0.340 & 0.018 & F & $10^{-2}$ & 0.131 & 0.219 & 0.019 & F \\ \hline
0.5 & 3.0 & $10^{-6}$ & 0.30  & 0.411 & 0.015  & F & $10^{-4}$ & 0.183 & 0.308 & 0.021 & F & $10^{-2}$ & 0.093 & 0.172 & 0.021 & F \\ \hline
0.5 & 3.2 & $10^{-6}$ & 0.315 & 0.424 & 0.017 & F & $10^{-4}$ & 0.19 & 0.296 & 0.023 & F & $10^{-2}$ & 0.096 & 0.149 & 0.023 & F \\ \hline
0.5 & 3.5 & $10^{-6}$ & 0.30 & 0.406 & 0.021 & F & $10^{-4}$ & 0.165 & 0.254 & 0.026 & F & $10^{-2}$ & 0.06 & 0.099 & 0.026 & F \\ \hline \hline 
1.0 & 2.5 & $10^{-6}$ & 0.31 & 0.464 & 0.015 & F & $10^{-4}$ & 0.25 & 0.479 & 0.029 & F & $10^{-2}$ & 0.21 & 0.503 & 0.047 & F \\ \hline
1.0 & 2.8 & $10^{-6}$ & 0.307 & 0.448 & 0.014  & F & $10^{-4}$ & 0.24 & 0.458 & 0.027 & F & $10^{-2}$ & 0.19 & 0.48 & 0.038 & F \\ \hline
1.0 & 3.0 & $10^{-6}$ & 0.313  & 0.44 & 0.012 & F & $10^{-4}$ & 0.23 & 0.44 & 0.027 & F & $10^{-2}$ & 0.18 & 0.467 & 0.04 & F \\ \hline
1.0 & 3.2 & $10^{-6}$ &  0.338 & 0.459 & 0.011  & F & $10^{-4}$ & 0.25 & 0.44 & 0.028 & F & $10^{-2}$ & 0.195 & 0.46 & 0.053 & F \\ \hline
1.0 & 3.5 & $10^{-6}$ & 0.33 & 0.447 & 0.011 & F & $10^{-4}$ & 0.27 & 0.427 & 0.028 & F & $10^{-2}$ & 0.2 & 0.44 & 0.071 & F \\
\bottomrule 
\end{tabular}
}
\label{P-V-constant-beta}
\end{table}
%%%%%%%%%%%%%%%%%%%%%%%%%%%%%%%%%%%%%%%%%

%%%%%%%%%%%%%%%%%%%%%%%%%%%%%%%%%%%%%%%%%
\begin{table}
\caption{Table of 230 GHz polarimetric quantities $\left( |m|_{\mathrm{net}}, \langle |m| \rangle,
|v_{\mathrm{net}}| \right) $ for different Sgr A* RIAF models. 
In each block of 4 rows/5 columns, we fix the positron fraction $f_{\mathrm{pos}} \in\{0.0, 0.01, 0.1\}$ and vary the plasma-$\beta\in\{10^{-2},10^{-1},1,10\}$. All other parameters are fixed to their values from \citet{Broderick2016}.
}
\makebox[ 0.83 \textwidth][c]{
\begin{tabular}{c|cccc|cccc|cccc}
\toprule
\hline
$\beta$ & $f_{\mathrm{pos}}$ & $|m|_{\mathrm{net}}$ & $\langle |m| \rangle $ & $ |v_{\mathrm{net}}| $ &  $f_{\mathrm{pos}}$ & $|m|_{\mathrm{net}}$ & $\langle |m| \rangle $ & 
$ |v_{\mathrm{net}}| $ &  $f_{\mathrm{pos}}$ & $|m|_{\mathrm{net}}$ & $\langle |m| \rangle $ & 
$|v_{\mathrm{net}}|$ 
\\ \midrule
$10^{-2}$ & 0.0 & 0.20 & 0.41 & 0.009 &   0.01 &  0.2 & 0.41 &  0.010 &  0.1 & 0.20 & 0.41 & 0.01 \\ \hline
$10^{-1}$ & 0.0 & 0.19 & 0.40 & 0.019 &   0.01 &  0.18 & 0.40 & 0.019 &   0.1 & 0.18 & 0.40 & 0.019  \\ \hline
$10^{0}$ & 0.0 & 0.18 & 0.39 & 0.032 &   0.01 & 0.18 & 0.39 & 0.032 & 0.1 & 0.18 & 0.39 & 0.033  \\ \hline
$10^{1}$ & 0.0 & 0.22 & 0.37 & 0.047 &  0.01 & 0.22 & 0.37 & 0.048 &  0.1 & 0.21 & 0.37 & 0.051  \\ 
\hline \midrule \hline
$10^{-2}$ & 0.5 & 0.2 & 0.40 & 0.015  & 1.0 & 0.20 & 0.40 & 0.017  &-&-&-&- \\ \hline
$10^{-1}$ & 0.5 & 0.18 & 0.40 & 0.023 & 1.0 & 0.19 & 0.39 & 0.027  &-&-&-&- \\ \hline
$10^{0}$ & 0.5 & 0.17 & 0.36 & 0.045 & 1.0 & 0.18 & 0.37 & 0.041  &-&-&-&- \\ \hline
$10^{1}$ & 0.5 & 0.17 & 0.37 & 0.054 & 1.0 & 0.17 & 0.36 & 0.058 &-&-&-&-  \\ 
\bottomrule 

\bottomrule 
\end{tabular}}
\label{P-V-RIAF}
\end{table}
%%%%%%%%%%%%%%%%%%%%%%%%%%%%%%%%%%%%%%%%%

%%%%%%%%%%%%%%%%%%%%%%%%%%%%%%%%%%%%%%%%%%%%%%%%%%%%%
\begin{figure*}
\center
\includegraphics[width=1.0\textwidth]{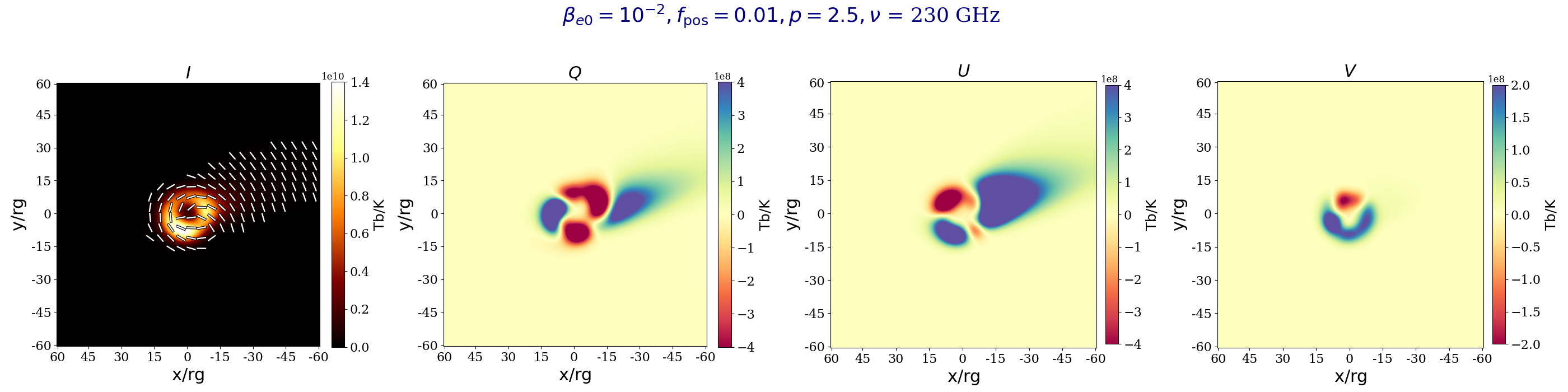}
\includegraphics[width=1.0\textwidth]{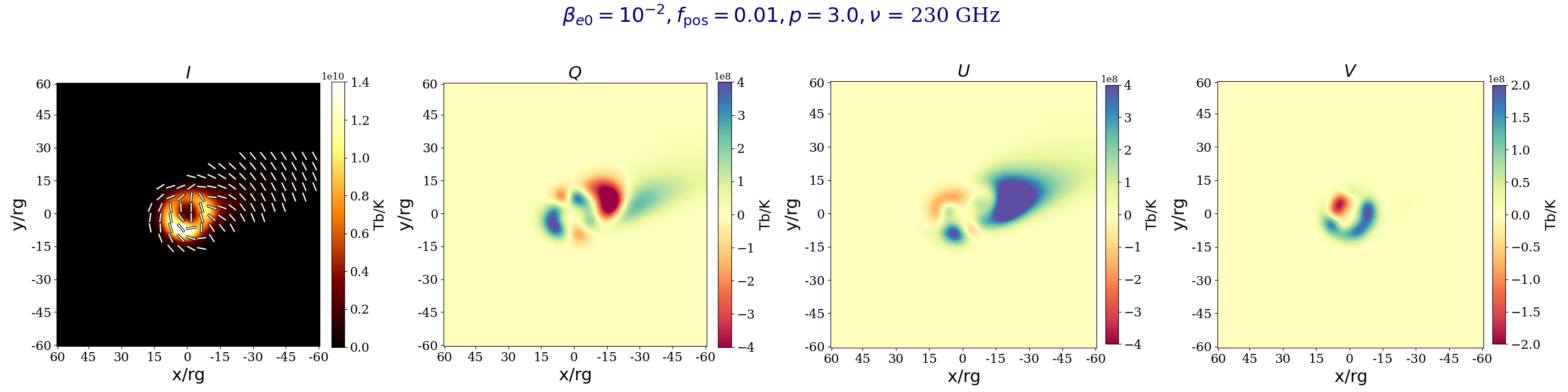}
\includegraphics[width=1.0\textwidth]{COMBINEED_Plane_1p00e-02_pos_0p01_pnth_3p5.png}
\caption{Stokes Map for M87* jet model with 
$\beta_{\mathrm{e0}}= 10^{-2}$ and for $p \in\{2.5, 3.0, 3.5\}$. We have chosen  $f_\mathrm{pos}= 0.01$. To facilitate the comparison between different models, we have chosen the same color bars for different models. All of the images are blurred with a fiducial EHT beam of $20\mu$as.}
\label{Constant-beta-1eM2-PNTH253}
\end{figure*}
%%%%%%%%%%%%%%%%%%%%%%%%%%%%%%%%%%%%%%%%%%%%%%%%%%%%%

%%%%%%%%%%%%%%%%%%%%%%%%%%%%%%%%%%%%%%%%%%%%
\begin{figure*}
\center
\includegraphics[width=0.99\textwidth]{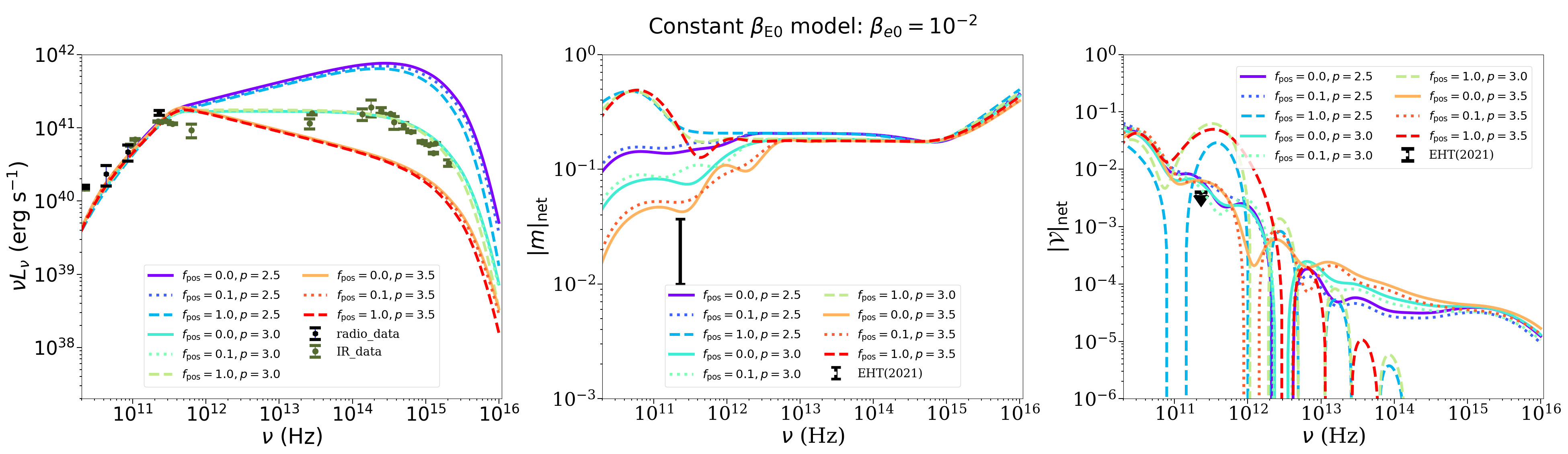}
\caption{ 
Polarized spectral decomposition for M87* jet models with electron spectral index $p \in\{2.5,3.0,3.5\}$ and positron-to-electron ratio $f_\mathrm{pos}\in\{0.0,0.1,1.0\}$. We fix  $\beta_{\mathrm{e0}}=10^{-2}$.}
\label{fig:PolSpectralFluxBetaE010Em2Pnth2Pt5And3And3Pt5Theta20fPos0AndPt5And1}
\end{figure*}
%%%%%%%%%%%%%%%%%%%%%%%%%%%%%%%%%%%%%%%%%%%% 

\section{Additional model exploration for M87* constant $\beta_{\rm e0}$ jet model}
\label{Beta-model-parameter-Exp}
Here we study the impact of changing the non-thermal index $p$ and the electron-to-magnetic pressure ratio $\beta_{\rm e0}$ on the Stokes maps and polarized SEDs from the M87* jet model.
In Fig. \ref{Constant-beta-1eM2-PNTH253}, we show images at a fixed value of $\beta_{\rm e0}=10^{-2}$ and $f_{\rm pos}=10^{-2}$ and vary $p\in{2.5, 3.0, 3.5}$. We show that increasing $p$ diminished the total linear and circular polarized intensity at 230 GHz. 
In Figure \ref{fig:PolSpectralFluxBetaE010Em2Pnth2Pt5And3And3Pt5Theta20fPos0AndPt5And1}, we present the total intensity SED and spectra of $|m|_{\rm net}$ and $|v|_{\rm net}$ when we vary the lepton spectral index and the positron fraction $(p, f_{\mathrm{pos}})$ in the range, $p \in\{2.5, 3.0, 3.5 \}$ and $ f_{\mathrm{pos}} \in\{0.0, 0.5, 1.0 \}$
while fixing $\beta_{\mathrm{e0}} = 10^{-2}$. 
It is evident that the total intensity Stokes $I$ SED is relatively insensitive to the positron fraction but it is highly sensitive to the electron spectral index $p$, as we would expect. Only $p=3.0$ fits the near-infrared data points in the M87* SED well. However, we have chosen $p=3.5$ as our fiducial M87* model because lower values of $p$ dramatically overproduce the observed linear and circular polarization (seen in the center and middle panels).  
In the linear polarization fraction $|m|$, the impact of different positron fractions $f_{\mathrm{pos}}$ is evident in lower frequencies $ \nu/\mathrm{Hz} \leq 10^{12}$; in particular, at $f_{\rm pos}=1$, all values of $p$ produce strong linear polarization fractions $\sim 30-40$\%. Finally, the spectrum of circular polarization $|v|$ shows a very distinctive dependence on the positron fraction $f_{\mathrm{pos}}$. At high frequencies $>10^{13}$Hz, increasing the positron fraction suppresses the Stokes V emission, as Faraday conversion is inefficient. However, near the 230 GHz spectral peak, the Stokes $V$ emission is dominated by Faraday conversion, and increasing the positron content increases the total flux density in the submm.

%%%%%%%%%%%%%%%%%%%%%%%%%%%%%%%%%%%%%%
\begin{figure*}
\center
\includegraphics[width=1.0\textwidth]{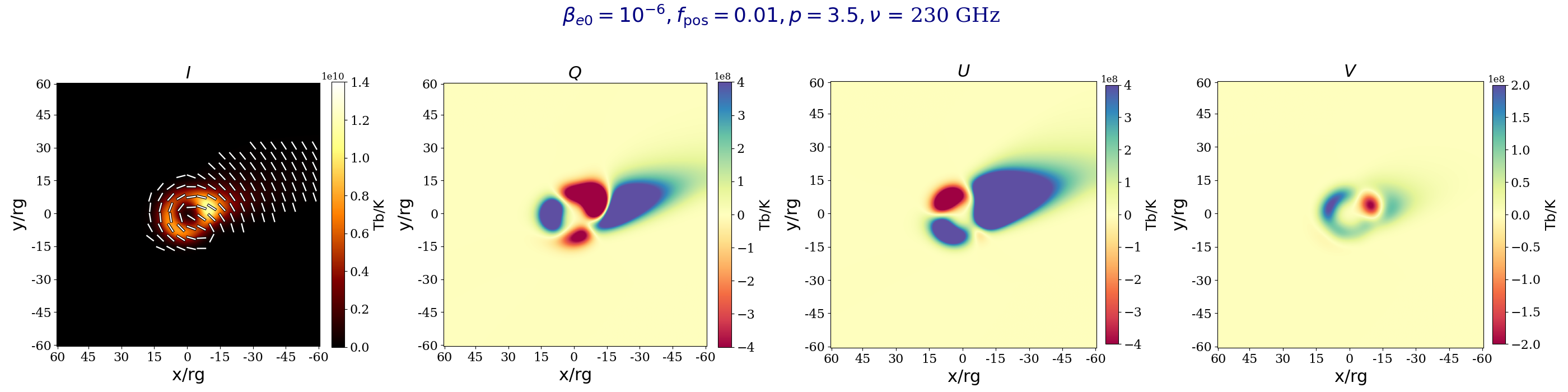}
\includegraphics[width=1.0\textwidth]{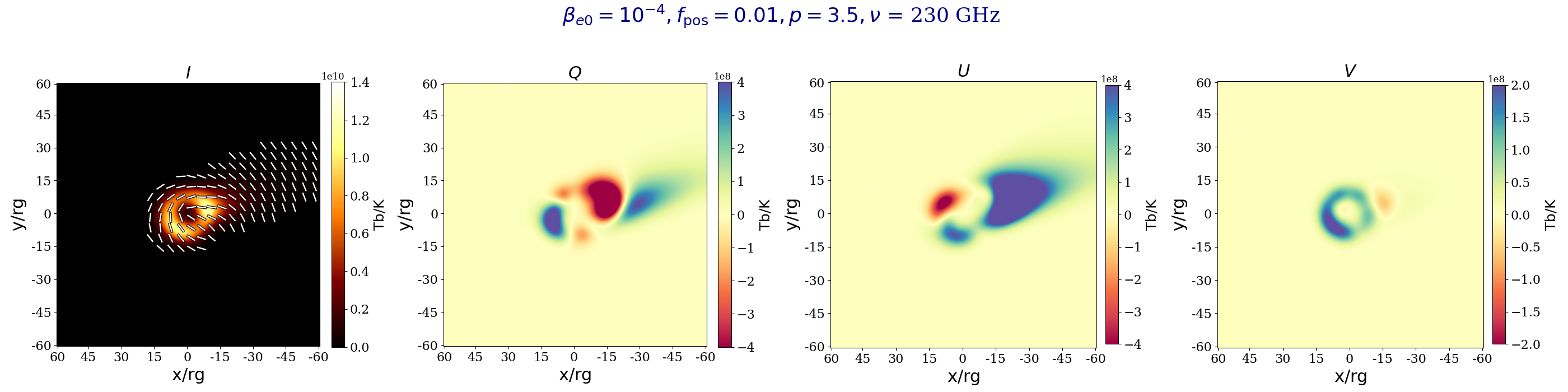}
\includegraphics[width=1.0\textwidth]{COMBINEED_Plane_1p00e-02_pos_0p01_pnth_3p5.png}
\caption{Stokes maps for M87* jet models with $\beta_{\mathrm{e0}}= (10^{-6}, 10^{-4}, 10^{-2})$, fixing  $p = 3.5$ and $f_{\rm pos}=0.01$. All of the images are blurred with a fiducial EHT beam of $20\mu$as.}
\label{Constant-beta-1eM642}
\end{figure*}
%%%%%%%%%%%%%%%%%%%%%%%%%%%%%%%%%%%%%%
\begin{figure*}
\includegraphics[width=0.99\textwidth]{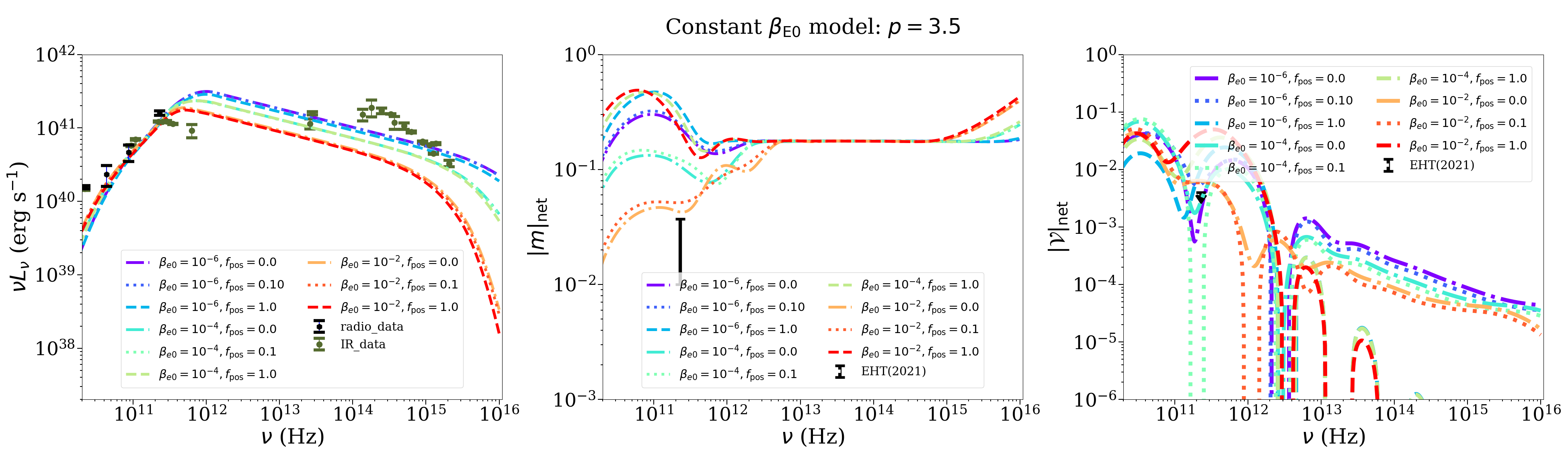}
\caption{
Polarized spectral decomposition for M87* jet models when varying $\beta_{\mathrm{e0}}\in\{10^{-6},10^{-4},10^{-2}\}$ and $f_\mathrm{pos}\in\{0.0,0.1,1.0\}$. We vary fix the electron distribution index $p=3.5$.}
\label{fig:PolSpectralFluxBetaE010Em6And10Em4And10Em2Pnth3Pt5Theta20fPos0AndPt5And1}
\end{figure*}
%%%%%%%%%%%%%%%%%%%%%%%%%%%%%%%%%%%%%%
In Fig. \ref{Constant-beta-1eM642}, we present Stokes maps varying $\beta_{\mathrm{e0}}\in\{10^{-6}, 10^{-4}, 10^{-2}\}$ while holding fixed $p=3.2$ and $f_{\mathrm{pos}} = 0.01$. 
We see that by increasing the electron pressure through increasing $\beta_{\mathrm{e0}}$, we 
decrease the polarized brightness in $Q$ and $U$. This arises because increasing $\beta_{\rm e0}$ increases the efficiency of Faraday rotation, as we can see by noticing more scrambling in the EVPA patterns as we proceed from the top to bottom in Fig. \ref{Constant-beta-1eM642}.
In Fig. \ref{fig:PolSpectralFluxBetaE010Em6And10Em4And10Em2Pnth3Pt5Theta20fPos0AndPt5And1}, we show Stokes $I$ SEDs and polarization fraction spectra when varying $\beta_{\rm e0}$. The $\beta_{\mathrm{e0}}=10^{-6}, 10^{-4}$ models tend to be more sensitive to $\beta_{\mathrm{e0}}$ than to positron content, with an X-ray excess (relative to the observed value) occurring for the lower value and the better fit to the $I$ spectrum occurring for the higher values. The polarized decomposition in  Fig. \ref{fig:PolSpectralFluxBetaE010Em6And10Em4And10Em2Pnth3Pt5Theta20fPos0AndPt5And1}, shows a pronounced effect for circular polarization, which may serve as a crucial observational discriminant with higher positron fraction drastically reducing the high frequency slope.
This is because Faraday conversion becomes less efficient at higher frequencies (note Faraday conversion is always efficient at the peak even when $f_\mathrm{pos}=1$).

In Figure \ref{fig:PolSpectralFluxBetaE010Em6And10Em4And10Em2Pnth3Pt5Theta20fPos0AndPt5And1}, we present the spectrum for $(\beta_{\mathrm{e0}}, f_{\mathrm{pos}})$ in the range, 
$\beta_{\mathrm{e0}} \in\{10^{-6}, 10^{-4}, 10^{-2} \}$ and $f_{\mathrm{pos}} \in\{0.0,  0.5, 1.0 \}$
keeping $p = 3.5$ fixed. It is evident that while the Stokes $I$ SED for different models is similar at lower frequencies where self-absorption dominates the spectral slope, smaller values of $\beta_{\rm e0}$ shift the position of the synchrotron peak and the position of the spectral cutoff at high $\nu$ to larger values. Models with larger $\beta_{\rm e0}$ (weaker magnetic fields), have their synchrotron emission peak at lower frequencies and also have a cutoff in the power-law slope from optically thin emission at lower frequencies (note that the upper Lorentz factor $\gamma_{\rm max}$ is fixed in all models). In the linearly polarized spectrum, the magnitude of the high frequency, near-infrared fractional linear polarization is constant with different $\beta_{\rm e0}$, but at submm and radio frequencies, models with larger $\beta_{\rm e0}$ values are less polarized. Once again, increasing $f_{\rm pos}$ increases the linear polarization fraction at all values of $\beta_{\rm e0}$ as a result of the suppression of Faraday rotation with increasing positron content \citep[e.g.][]{Jimenez2018}. Particularly for low $\beta_{\rm e0}$ at high frequencies in Stokes $I$ or $m$, it is easier to distinguish models based on $p$ or $\beta_{e0}$ than with positron fraction $f_\mathrm{pos}$, but at lower sub-mm and radio frequencies,  $f_\mathrm{pos}$ has the dominant effect. The spectrum of the circular polarization behaves similarly as in Figure \ref{fig:PolSpectralFluxBetaE010Em2Pnth2Pt5And3And3Pt5Theta20fPos0AndPt5And1}. The circular polarization fraction $|V|$ diminishes with higher positron fraction as the Faraday conversion becomes less efficient at high frequencies and since the intrinsic production of Stokes $|V|$ is suppressed due to the presence of positrons in the emitting plasma. However, Faraday conversion is always efficient at the peak even when $f_\mathrm{pos}=1$ and the pair plasma models have the highest degree of circular polarization at submm wavelengths as their Faraday conversion coefficients are enhanced. 
%%%%%%%%%%%%%%%%%%%%%%%%%%%%%%%%
\section{Impact of Faraday Rotation and conversion}
\label{Faradays}
\begin{figure*}
\center
\includegraphics[width=1.0\textwidth]{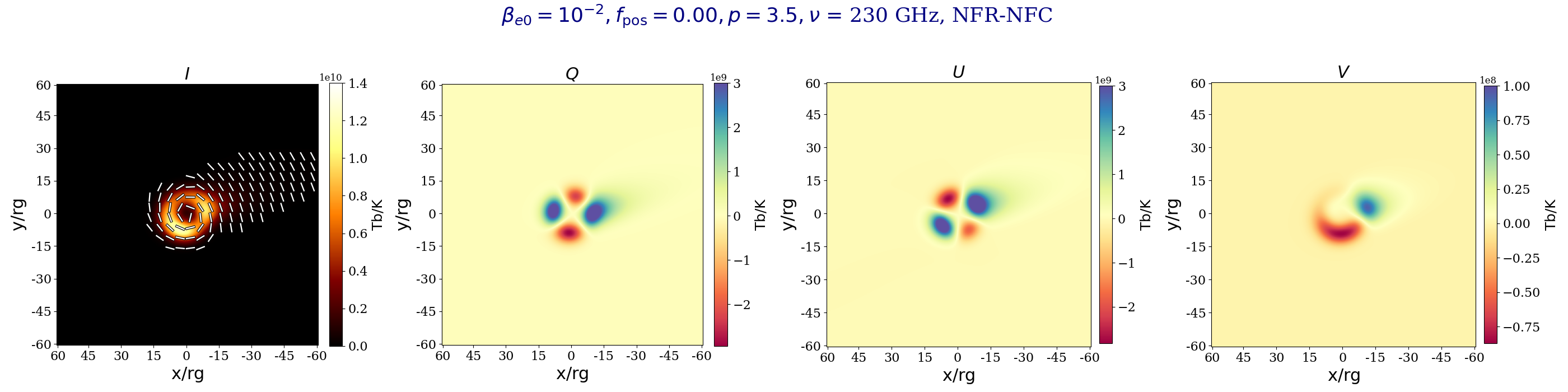}
\includegraphics[width=1.0\textwidth]{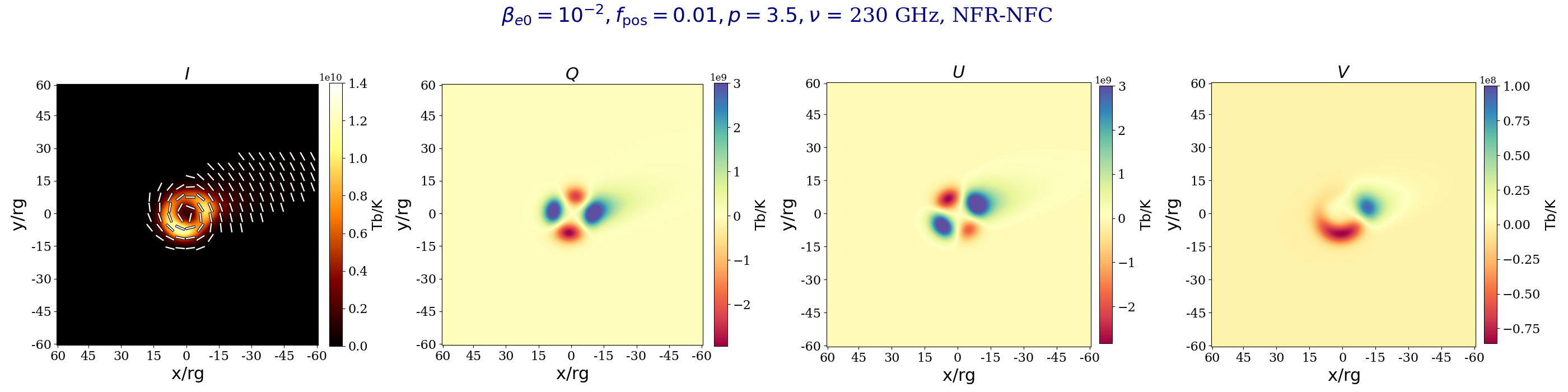}
\includegraphics[width=1.0\textwidth]{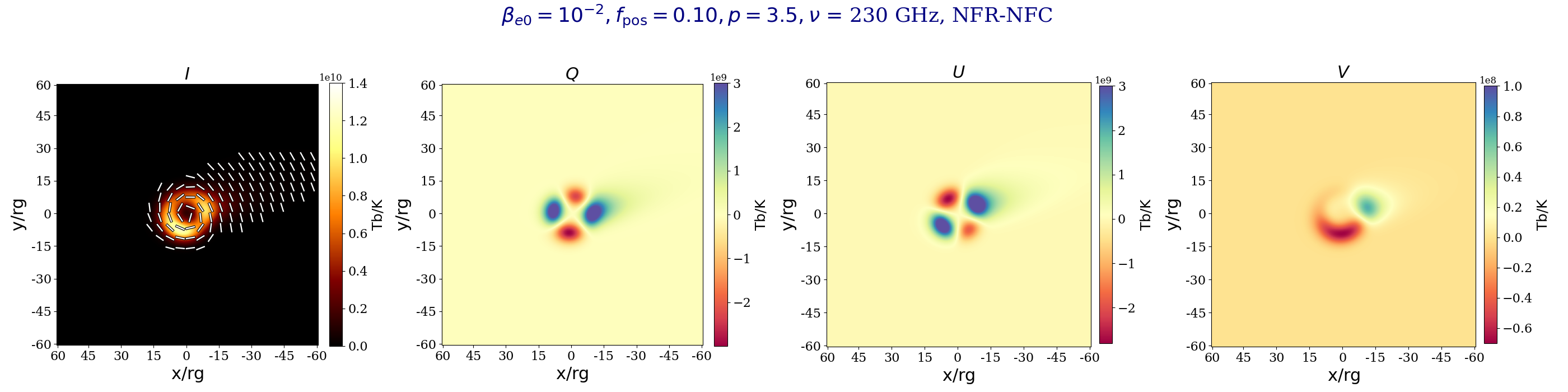}
\includegraphics[width=1.0\textwidth]{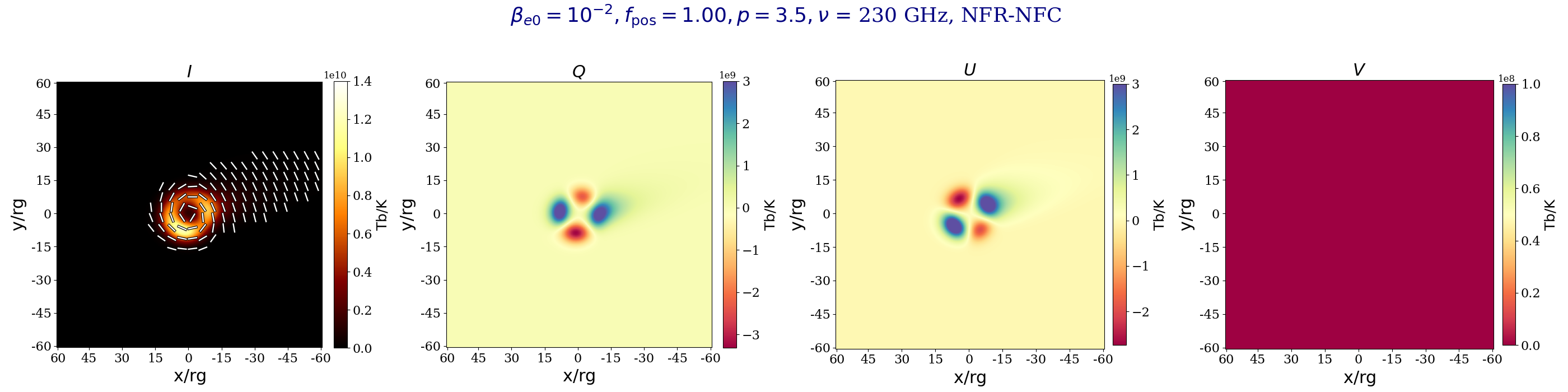}
\caption{Stokes maps for M87* jet models were we fix $\beta_{\mathrm{e0}}=10^{-2}$ and $p$ = 3.5. From top to bottom, we vary the fraction of positrons $f_{\rm pos}\in\{0, 0.01, 0.1, 1\}$. Here we have turned off the Faraday rotation and Faraday conversion coefficients in performing radiative transfer.All of the images are blurred with a fiducial EHT beam of $20\mu$as.} \label{StokesMapsBetaE010Em2Pnth3Pt2Theta20fPos0AndPt5And1_NFR}
\end{figure*}

In this Appendix, we demonstrate the dominant role of Faraday rotation and conversion in determining the submm linear and circular polarization structure of our models. In Figure 
\ref{StokesMapsBetaE010Em2Pnth3Pt2Theta20fPos0AndPt5And1_NFR},
we present images of our fiducial model (varying the positron fraction $f_{\rm pos}$ in each row) with both Faraday conversion and rotation manually set to zero in our radiative transfer. Comparing to Figures~\ref{StokesMapsBetaE010Em2Pnth3Pt2Theta20fPos0AndPt5And1} and \ref{StokesMapsBetaE010Em2Pnth3Pt2Theta20fPos0AndPt5And1}, we find the most significant changes in the model images when turning on and off Faraday effects are:  

$\bullet$ With Faraday rotation turned off, the linear polarization EVPA pattern at all values of $f_{\rm pos}$ is entirely azimuthal (and the Stokes $Q$ and $U$ patterns are completely quadrupolar). This is because in the absence of Faraday effects, the EVPA orientation is only set by the field orientation in the emission region, which in our models near the horizon is predominantly poloidal. Faraday rotation scrambles the EVPA pattern, depolarizing the image on EHT scales and distorting the underlying azimuthal pattern that is the imprint of the jet launching field structure in this model. 

$\bullet$ With Faraday conversion turned off, the circular polarization $V$ diminishes with increasing $f_{\rm pos}$ and completely disappears when $f_\mathrm{pos}\to 1$. Circular polarization at $230$GHz in these models is predominantly due to Faraday conversion. With Faraday conversion present, a large positron fraction actually enhances the total degree of circular polarization.  

%%%%%%%%%%%%%%%%%%%%%%%%%%%%%%%%%%%%%%%%%%%%%%%%%%%%
%%%%%%%%%%%%%%%%%%%%%%%%%%%%%%%%%%%%%%%%%%%%%%%%%%

\end{document}